\documentclass[12pt]{article}  %
\newenvironment{keywords}{\textbf{Keywords:}\begin{center}}{\end{center}}

\usepackage[a4paper,hscale=.7,vscale=.7,centering]{geometry} %

\IfFileExists{srcltx.sty}{\usepackage[active]{srcltx}} %

\usepackage{amsmath,amssymb} %
\usepackage{xspace,bm,paralist,graphicx,color}
\usepackage[normalem]{ulem} %
\newcommand{\rmi}[1]{{\mbox{\scriptsize #1}}}%
\def\lsi{\raise0.3ex\hbox{$<$\kern-0.75em\raise-1.1ex\hbox{$\sim$}}}
\def\gsi{\raise0.3ex\hbox{$>$\kern-0.75em\raise-1.1ex\hbox{$\sim$}}}
\newcommand{\lsim}{\mathop{\;\lsi\;}}
\newcommand{\gsim}{\mathop{\;\gsi\;}}

\newcommand{\CH}{\mathcal{H}}

\newcommand{\GeV}{\ensuremath{\;\mathrm{GeV}}} %
\newcommand{\gev}{\ensuremath{\;\mathrm{GeV}}} %
\newcommand{\mev}{\ensuremath{\;\mathrm{MeV}}} %
\newcommand{\kev}{\ensuremath{\;\mathrm{keV}}} %
\newcommand{\ev} {\ensuremath{\;\mathrm{eV}}}   %
\newcommand{\mpc}{\;\mathrm{Mpc}} 
\newcommand{\dm}{{\textsc{dm}}} 
\newcommand{\numsm}{$\nu$MSM\xspace} 
\newcommand{\lya}{Lyman-$\alpha$\xspace} 
\newcommand{\rp}{\textsc{rp}} %
\newcommand{\cdm}{\textsc{cdm}\xspace} %
\newcommand{\wdm}{\textsc{wdm}\xspace} %
\newcommand{\fnrp}{{\ensuremath{f_\textsc{nrp}}}\xspace}%

\usepackage[square,comma]{natbib} %
\usepackage{hyperref} %

\IfFileExists{hypernat.sty}{\usepackage{hypernat}}

\begin{document}
\title{The Role of Sterile Neutrinos in Cosmology and Astrophysics}
\author{Alexey Boyarsky\thanks{Institute of Theoretical Physics, \small ETH
    H\"{o}nggerberg, CH-8093 Z\"urich, Switzerland} \thanks{Bogolyubov
    Institute for Theoretical Physics, Kiev 03680,Ukraine}~,~ Oleg
  Ruchayskiy\thanks{Institute of Theoretical Physics \'Ecole Polytechnique
    F\'ed\'erale de Lausanne, FSB/ITP/LPPC, BSP, CH-1015, Lausanne,
    Switzerland}~~and Mikhail Shaposhnikov$^\ddagger$} \date{}

\markboth{Boyarsky, Ruchayskiy \& Shaposhnikov}%
{Sterile neutrinos in cosmology and astrophysics}

\maketitle

\begin{abstract}
  We present a comprehensive overview of an extension of the Standard Model
  that contains three right-handed (sterile) neutrinos with masses below the
  electroweak scale [the \emph{Neutrino Minimal Standard Model}, (\numsm)]. We
  consider the history of the Universe from the inflationary era through today
  and demonstrate that most of the observed phenomena \emph{beyond the
    Standard Model} can be explained within the framework of this model. We
  review the mechanism of baryon asymmetry of the Universe in the \numsm and
  discuss a dark matter candidate that can be warm or cold and satisfies all
  existing constraints. From the viewpoint of particle physics the model
  provides an explanation for neutrino flavor oscillations.  Verification of
  the \numsm is possible with existing experimental techniques.
\end{abstract}


\section{Introduction}
\label{se:intro}
The Standard Model (SM) of elementary
particles~\cite{Glashow:1961tr,Weinberg:1967tq,Salam:1968rm}, defined as a
renormalizable field theory, based on the SU(3)$\times$SU(2)$\times$U(1) gauge
group, and containing three fermionic families --- left-handed particles,
SU(2) doublets, right-handed particles, SU(2) singlets (no right-handed
neutrinos) and one Higgs doublet --- has successfully predicted a number of
particles and their properties.  However, there is no doubt that the SM is not
a final theory.  Indeed, over the past several decades it has become
increasingly clear that it fails to explain a number of \emph{observed}
phenomena in particle physics, astrophysics, and cosmology. These phenomena
\emph{beyond the SM} (BSM) are
\begin{inparaenum}[\sl (a)]
\item neutrino oscillations (transition between neutrinos of different
  flavors),
\item baryon asymmetry (excess of matter over anti-matter in the
  Universe),
\item dark matter (about 80\% of all matter in the Universe consisting of
  unknown particles),
\item inflation
  (a period of  rapid accelerated expansion in the early Universe), and
\item {dark energy} (late-time accelerated expansion of the Universe)
\end{inparaenum}
This list of {\em well-established observational} drawbacks of the SM is
considered complete at present. All the other BSM problems --- for, instance,
the gauge hierarchy and strong-CP problems --- require theoretical
fine-tuning.

At what energies should the SM be superseded by some other, more fundamental
theory?  The existence of gravity with the coupling related to the Planck
scale --- $M_{Pl}=G_N^{-1/2}=1.2\times 10^{19}$ GeV, where $G_N$ is the
Newtonian gravitational constant --- implies that the cut-off is \emph{at
  least} below the Planck scale. If the cutoff is identified with $M_{Pl}$,
the low-energy Lagrangian can contain all sorts of higher-dimensional
SU(3)$\times$SU(2)$\times$U(1)-invariant operators that are suppressed by the Planck
scale,
\begin{equation}
  \mathcal{L} = \mathcal{L}_{\rm SM}
  +\sum_{n=5}^\infty\frac{\mathcal{O}_n}{M_{Pl}^{n-4}}~,
\label{lagr}
\end{equation}
where $\mathcal{L}_{SM}$ is the Lagrangian of the SM.  These operators lead to
a number of physical effects that cannot be described by the SM, such as
neutrino masses and mixings, proton decay, etc. However, as we will discuss
below, even the theory shown in Eq.~(\ref{lagr}) does not survive when
confronted with different experiments and observations.

Alternatively, one can place a cut-off $\Lambda \ll M_{Pl}$ in
Eq.~(\ref{lagr}), which would imply that new physics (and new particles)
appears well below the Planck scale at energies $E\sim \Lambda$. If $\Lambda
\gg M_W$, where $M_W$ is the mass of the weak $W$ boson, the resulting theory
suffers from the so-called \emph{gauge hierarchy problem}, that is, the
problem of quantum stability of the mass of the Higgs boson against quantum
corrections from heavy particles.

Most of the research in BSM physics carried out during the past few decades
was devoted to solving the gauge hierarchy problem.  Many different
suggestions were proposed concerning how to achieve the ``naturalness'' of
electroweak symmetry breaking.  These propositions are based on supersymmetry,
technicolor, and large extra dimensions, among other ideas. Finding a solution
to the gauge hierarchy problem, coupled with the need to solve observational
and other fine-tuning problems of the SM, is extremely challenging.  Most of
the approaches postulate the existence of new particles with masses
\emph{above} the electroweak scale (ranging from $10^2$~GeV to
$10^{15}$--$10^{16}$~GeV).  As a result, the proposed theories contain a
plethora of (not yet observed) new particles and parameters.

In this review, we describe  a conceptually different scenario
for BSM physics and its consequences for astrophysics and cosmology in
an attempt to address the BSM problems named above \emph{without} introducing
new energy scales (that is, in addition to the electroweak and the Planck
scales).  In such an approach, the hierarchy problem is shifted to the Planck
scale, and there is no reason to believe that the field theoretical logic
is still applicable to it.

Below we show (following Refs.~\cite{Asaka:05a,Asaka:05b} and a number of
subsequent works) that this goal may be achieved with a very simple extension
of the SM.  The only new particles, added to the SM Lagrangian are three
gauge-singlet fermions (i.e., \emph{sterile neutrinos}) with masses
\emph{below} the electroweak scale.  Right-handed neutrinos are strongly
motivated by the observation of neutrino flavor
oscillations. In Section~\ref{sec:neutr-oscill} we  review  neutrino oscillations and
introduce the corresponding Lagrangian.
We summarize the choice of parameters of the Neutrino Minimal Standard Model
(\numsm) in Section~\ref{sec:phenomenology-numsm}. In
Section~\ref{sec:cosm-early-univ}, we present a \numsm cosmology. We discuss
the restrictions from astrophysics, cosmology, and particle physics
experiments, as well as future searches in Section~\ref{sec:testing-numsm}.
In Section ~\ref{sec:beyond}, we conclude with a discussion of possible
extensions of the \numsm and potential astrophysical applications of sterile
neutrinos.

\section{Neutrino oscillations}
\label{sec:neutr-oscill}
When the SM was conceived, neutrinos were thought to be massless and different
lepton numbers were believed to be conserved (which would not be the case if
right-handed neutrinos were present). This was a reason for {\em not
  introducing right-handed neutrinos} decades ago.

Transitions between neutrinos of different flavors have been observed. These
\emph{neutrino oscillations} manifest themselves in experiments with solar,
atmospheric, accelerator and reactor neutrinos.  The subject of neutrino
oscillations has received a great deal of attention in recent years and many
reviews are available (see e.g.  Refs.~\cite{Strumia:06,Giunti:06}).  The
observed oscillations determine the mass differences between propagation
states~\cite{Schwetz:08a}
\begin{equation}
  \label{eq:1}
  \Delta
  m^2_\mathrm{sun}=7.65^{+0.23}_{-0.20}\times 10^{-5} \ev^2
\end{equation}
and
\begin{equation}
  \label{eq:2}
  |\Delta m^2_\text{atm}| = 2.40^{+0.12}_{-0.11}\times 10^{-3}\ev^2.
\end{equation}
Importantly, each of these results was confirmed by two different types of
experiments: accelerator~\cite{MINOS:08} for Eq.~(\ref{eq:2}) and
reactor~\cite{KamLAND:08} for Eq.~(\ref{eq:1}).

The observed pattern of neutrino oscillations cannot be explained by
the action Eq.~(\ref{lagr}) with the Planck-scale cutoff. Indeed, the
lowest-order five-dimensional operator
\begin{equation}
  \mathcal{O}_5 = A_{\alpha\beta}
  \left({\bar L_\alpha}\tilde\phi\right)
  \left(\phi^\dagger L_\beta^c\right)
\label{dim5}
\end{equation}
leads to the Majorana neutrino masses of the order $m_\nu\sim v^2/M_{Pl}
\simeq 10^{-6}$ eV, where $L_\alpha$ are left-handed leptonic doublets, the index
$\alpha=e,\mu,\tau$ labels generations, $\phi$ is a Higgs doublet with $\tilde
\phi_j = i(\tau_2)_{j}^k \phi^*_k$, $c$ is the sign of charge conjugation, and
$v=174$ GeV is the vacuum expectation value of the Higgs field.

The fact that the $m_\nu$ following from this Lagrangian is so small compared
with the lower bound on neutrino mass arising from the observations of
neutrino oscillations ($m_\nu > \sqrt{|\Delta m^2_{\rmi{atm}}|}\simeq
0.05$~eV) rules out the conjecture that the theory shown in Eq.~(\ref{lagr})
is a viable effective field theory up to the Planck scale. Therefore,
\emph{the existence of neutrino oscillations requires adding new particles to the Lagrangian
  of the Standard Model.}

Let us add $\mathcal{N}$ right-handed neutrinos $N_I$ ($I=1,\mathcal{N}$).
The most general renormalizable Lagrangian has the form:
\begin{eqnarray}
  \label{lagr1}
  \mathcal{L} = \mathcal{L}_{SM}+ i \bar N_I \partial_\mu \gamma^\mu N_I -
  \left(F_{\alpha I} \,\bar L_\alpha N_I \tilde \phi 
    - \frac{M_I}{2} \; \bar {N_I^c} N_I + h.c.\right)~,
\end{eqnarray}
where $F_{\alpha I}$ are new Yukawa couplings. The right-handed neutrinos have
zero electric, weak and strong charges; therefore, they are often termed
``singlet,'' or ``sterile'' fermions.  The Majorana masses $M_I$ are
consistent with the gauge symmetries of the SM.  Without loss of generality,
the Majorana mass matrix in diagonal form can be chosen.

If the Dirac masses $M_D= F_{\alpha I} \langle \phi \rangle$ are much smaller
than the Majorana masses $M_I$, \emph{the type I seesaw
  formula}~\cite{Minkowski:77,Ramond:79,Mohapatra:79,Yanagida:80} holds that
\begin{equation}
  (m_\nu)_{\alpha\beta} = -\sum_{I=1}^\mathcal{N}
  (M_D)_{\alpha I} \frac{1}{M_I} (M_D^T)_{I\beta},
\label{see-saw}
\end{equation}
where $m_\nu$ is a $3\times 3$ matrix of active neutrino masses, mixings, and
(possible) CP-violating phases.  An elementary analysis of Eq.~(\ref{see-saw})
shows that the number of right-handed singlet fermions $\mathcal{N}$ must be
\emph{at least two} to fit the data of neutrino oscillations.  If there were
only one sterile neutrino, then the two active neutrinos would be massless.
If there were two singlet fermions, only one of the active neutrinos would be
massless, which does not contradict the results from experiment.  Moreover, in
this case there are 11 new parameters in the Lagrangian~(\ref{lagr1}) --- two
Majorana masses, two Dirac masses, four mixing angles, and three CP-violating
phases --- which is more than the number of parameters (7) describing the mass
matrix of active neutrinos with one zero eigenvalue. In other words, for
${\cal N}=2$ the Lagrangian~(\ref{lagr1}) can describe the pattern of neutrino
masses and mixings observed experimentally. Of course, the situation is even
more relaxed for ${\cal N}= 3$.  The Lagrangian (\ref{lagr1}) with
$\mathcal{N}=3$ restores the symmetry between quarks and leptons: Every left
quark and lepton has a right counterpart.

The seesaw formula~(\ref{see-saw}) allows the mass of singlet neutrinos to be
a free parameter: Multiplying $M_D$ by any number $x$ and $M_I$ by $x^2$ does
not change the right-hand side of the formula. Therefore, the choice of $M_I$
is a matter of theoretical prejudice that cannot be fixed by active-neutrino
experiments alone.

In this review, our choice of $M_I$ is roughly of the order of the other mass
term in the Lagrangian of the SM, the mass of the Higgs boson.  This choice
does not lead to any intermediate scale between the electroweak and Planck
scales, but it does require small Yukawa couplings $F_{\alpha I}$. It allows
us to solve, in a unified manner, the observational problems of the SM
discussed above.  Specifically, the parameters of the model~(\ref{lagr1}) can
be chosen such that it provides a mechanism to generate baryon asymmetry of
the Universe (BAU) and a dark matter candidate in the form of the lightest
sterile neutrino. Also, this theory can accommodate inflation, if one
considers non-minimal coupling between the Higgs boson and
gravity~\cite{Bezrukov:07}. Finally, a scale-invariant extension of the
model~\cite{Shaposhnikov:08b,Shaposhnikov:08c} that includes unimodular
gravity may solve the problem of stability of the Higgs mass against radiative
corrections, even if the Planck scale is included, and could lead to an
explanation of dark energy and of the absence of the cosmological constant.

The Lagrangian~(\ref{lagr1}) with such a choice of parameters, explaining all
confirmed BSM phenomena,\footnote{There are also several unexplained phenomena
  in astrophysics (e.g., the 0.511-MeV line from the Galactic
  center~\cite{Weidenspointner:06}, pulsar-kick velocities~\cite{Lyne:94}, a
  feature in the positron spectrum at $\sim 100$~GeV~\cite{PAMELA:08a}, etc.).
  We do not consider them in this review, as for these phenomena standard
  physics explanations may exist.  Note, however, that the sterile neutrinos
  may play an important role in some of these cases (see
  Section~\ref{sec:beyond}). In addition, several anomalies in particle
  physics experiments (such as evidence for the neutrinoless double-beta
  decay~\cite{nu0bb} and the annual modulation of signal, observed by
  DAMA/LIBRA~\cite{DAMA:08}) have been reported.  However, none of them has
  yet been confirmed by other experiments.} %
has been termed the \numsm --- the \emph{Neutrino Minimal Standard
  Model}~\cite{Asaka:05b,Asaka:05a}.

\section{A choice of parameters: the \numsm}
\label{sec:phenomenology-numsm}
In this section we  discuss in  detail how the parameters of
the Lagrangian~(\ref{lagr1}) should be chosen so as to explain
the BSM phenomena without introducing  new physics above the
electroweak scale.

\subsection{Neutrino masses and oscillations}
\label{sec:neutr-mass}
As discussed above, the Lagrangian~(\ref{lagr1}) can explain any pattern of
active neutrino masses and mixings.  A naturalness
argument~\cite{Vissani:1997ys}, reproduced below, indicates that the masses
$M_I$ should be smaller than $10^{7}$ GeV.  Indeed, the stability of the Higgs
mass against radiative corrections, which arise from $N_I$-exchange in the
loops, requires
\begin{equation}\frac{1}{16\pi^2}\sum_{\alpha
  I}|F_{\alpha I}|^2 M_I^2 \lsim M_H^2~,
\label{Hcorr}
\end{equation}
where $M_H$ is the Higgs mass. Combining this equation with the seesaw formula
(\ref{see-saw}) leads to
\begin{equation}
  M_I \lsim \left(\frac{16\pi^2 M_H^2 v^2}{\sqrt{|\Delta
        m^2_\text{atm}|}}\right)^{\frac{1}{3}}
  \simeq 10^7 {\rm GeV}~.
\end{equation}
Any $M_I$ mass below this value is ``natural''. Also note that setting $M_I$
to zero increases the symmetry of the Lagrangian (introducing lepton number
conservation) and thus this choice is also stable against radiative
corrections. In the \numsm approach, we set $M_I$ below the electroweak scale,
so that these new particles are on the same footing as known quarks and
leptons.

The observed existence of two mass splittings (solar and atmospheric,
Eqs.~(\ref{eq:1}--\ref{eq:2})) requires at least two sterile neutrinos.  To
estimate what neutrino data imply for the Yukawa couplings, we take the larger
of the two mass splittings $|\Delta m^2_\text{atm}|$ and find from the seesaw
relations~(\ref{see-saw}) that
\begin{equation} 
  \label{eq:10}
  |F|^2 \approx
  \frac{\sqrt{|\Delta m_\textrm{atm}^2|} M_I}{v^2} \sim 2\times
  10^{-15}\frac{M_I}{\rm GeV}~.
\end{equation}
where $|F|^2$ is a typical value of Yukawa couplings $F_{\alpha I}$.  The
condition $M_I \lesssim 10^2\gev$ implies that $|F|^2\lesssim 10^{-13}$.
Clearly, the theory given in Eq.~(\ref{lagr1}) can describe neutrino
oscillation data even if one of the three sterile neutrinos has arbitrarily
small Yukawa couplings.

Another possible choice of parameters of the Lagrangian~(\ref{lagr1}),
suffering from the hierarchy problem described in
Reference~\cite{Vissani:1997ys}, relates the smallness of the neutrino masses
to the largeness of the Majorana mass terms $M_I$. Indeed, if we take the
Yukawa couplings $F_{\alpha I}\sim 1$ and the Majorana masses in the range
$M_I \sim 10^{10}-10^{15} \GeV$, we obtain the neutrino masses, as required by
experimental observations.  An attractive feature of this scenario is that
this new scale may be associated with the Grand Unification Theory (GUT)
scale.  A model~(\ref{lagr1}) with this choice of parameters can also give
rise to BAU through leptogenesis \cite{Fukugita:1986hr} and anomalous
electroweak number non-conservation at high temperatures \cite{Kuzmin:1985mm}.
However, we do not discuss this possibility here; for review of the GUT-scale
seesaw and the thermal leptogenesis scenario associated with it, see, for
instance, Reference~\cite{Davidson:2008bu}.\footnote{There is still another
  choice of seesaw parameters with a sterile neutrino mass $\sim
  1$~eV~\cite{deGouvea:05}, which we do not discuss here.}

\subsection{Dark matter candidate}
\label{sec:dm}
It has been noticed that a sterile neutrino may make an interesting dark
matter
candidate~\cite{Dodelson:93,Shi:98,Dolgov:00,Abazajian:01a,Abazajian:01b}. 
In
the \numsm, a dark matter sterile neutrino is simply one of the singlet
fermions (for definiteness, we consider it to be $N_1$). The interaction
strength between the sterile neutrino and the matter is \emph{superweak} with the
characteristic strength $\theta\, G_F$, where $G_F$ is the Fermi constant and
where the \emph{mixing angle} $\theta\ll1$ is defined as
\begin{equation}
  \label{eq:3}
  \theta_1^2 = \sum_{\alpha = e,\mu,\tau} \frac{v^2|F_{\alpha 1}|^2}{M_1^2}\;.
\end{equation}
Based on the universal Tremaine--Gunn bound~\cite{Tremaine:79}, the masses of
sterile neutrinos are restricted to the keV range and above.  Specifically,
for fermionic DM particles an \emph{average} phase-space density in any dark
matter-dominated system cannot exceed the density given by the Pauli exclusion
principle.  Applied to the smallest dark matter-dominated objects [dwarf
spheroidal galaxies of the Milky Way, (dSphs)] this bound translates into
$M_\dm \ge 400 \ev$~\cite{Boyarsky:08a}.  

A sterile neutrino is an example of \emph{decaying} DM.  Through its mixing
with the ordinary neutrinos, $N_1$ can decay (via $Z$ boson exchange) into
three (anti)neu\-tri\-nos.  To be DM, the lifetime of $N_1$ should be greater
than the age of the Universe, which restricts the mixing angle. A
significantly stronger constraint comes from a subdominant one-loop decay
channel into a photon and an active neutrino.  The energy of the produced
photon is $E_\gamma= M_1/2$, and the decay width is given
by~\cite{Pal:81,Barger:95}
\begin{equation}
  \label{eq:4}
  \Gamma_{N_1\to\gamma\nu} = \frac{9\, \alpha\, 
    G_F^2} {1024\pi^4}\sin^2(2\theta_1)\, M_1^5  
  \simeq 5.5\times10^{-22}\;\theta_1^2
  \left[\frac{M_1}{\mathrm{keV}}\right]^5\;\mathrm{s}^{-1}\;,
\end{equation}
where $\alpha$ is the fine-structure constant.  The expression for the width
of the dominant decay channel has the same parametric form, with the numeric
coefficient being $\sim 128$ times larger.  This radiative decay would produce
a narrow line, for instance, in the diffuse X/$\gamma$-ray
background~\cite{Dolgov:00,Abazajian:01b}. This gives a restriction on the
mixing angle,
\begin{equation}
  \theta_1^2 \lesssim 1.8\times
10^{-5}\left(\frac{\mathrm{keV}}{M_1}\right)^5, 
\label{eq:14}
\end{equation}
meaning that the lifetime of sterile neutrino DM $\tau_{N_1\to3\nu}$ exceeds
$10^{24}$~sec~\cite{Boyarsky:08b} --- six orders of magnitude longer than
the age of the Universe. We  discuss X-ray constraints in detail in
Section~\ref{sec:restr-astro}.

The contribution of the $N_1$ sterile neutrino to the masses of the active
neutrinos is of the order $\delta m_\nu \sim\theta_1^2 M_1$.  The lifetime
constraints mean that for $M_1\gtrsim 2\kev$ the correction $\delta m_\nu$ is
less than the error bar in the solar neutrino mass difference~(\ref{eq:1}).
This implies that the neutrino $N_1$ cannot contribute significantly to the
active neutrino mass matrix. As we show below this conclusion is valid for all
admissible masses and mixing angles.

\emph{Thus, at least three sterile neutrinos are required to account for DM
  and explain neutrino oscillations}.  In spite of the very weak coupling of
DM sterile neutrino $N_1$ with matter, it could have been produced in
sufficient amounts in the early Universe (see
Section~\ref{sec:dm-production-numsm} below).

\subsection{Two heavier neutrinos}
\label{sec:two-heavy-nu}
Two other sterile neutrinos ($N_{2},N_3$) should interact with the Standard
Model particles more strongly than $N_1$ to explain the observed pattern of
neutrino oscillations.  Here we outline properties of $N_2$ and $N_3$ upon
which we elaborate further in Section~\ref{sec:cosm-early-univ}.

\begin{figure}
  \includegraphics[width=0.5\linewidth]{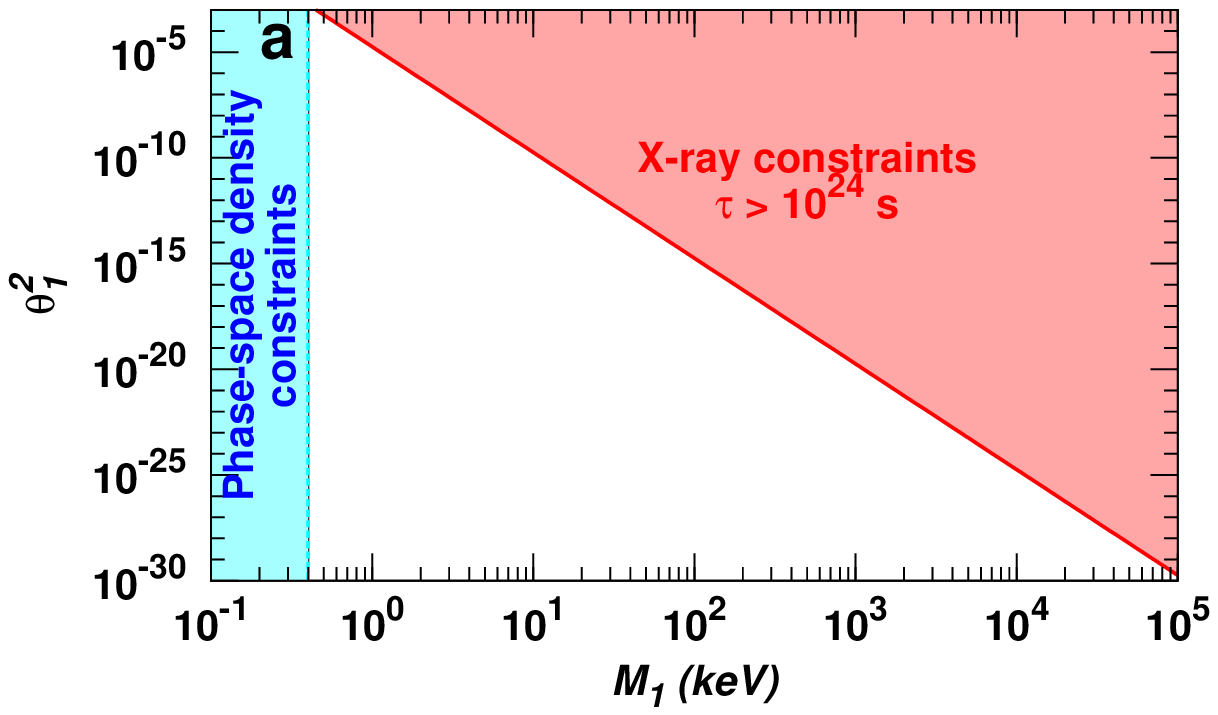}
  \includegraphics[width=0.5\linewidth]{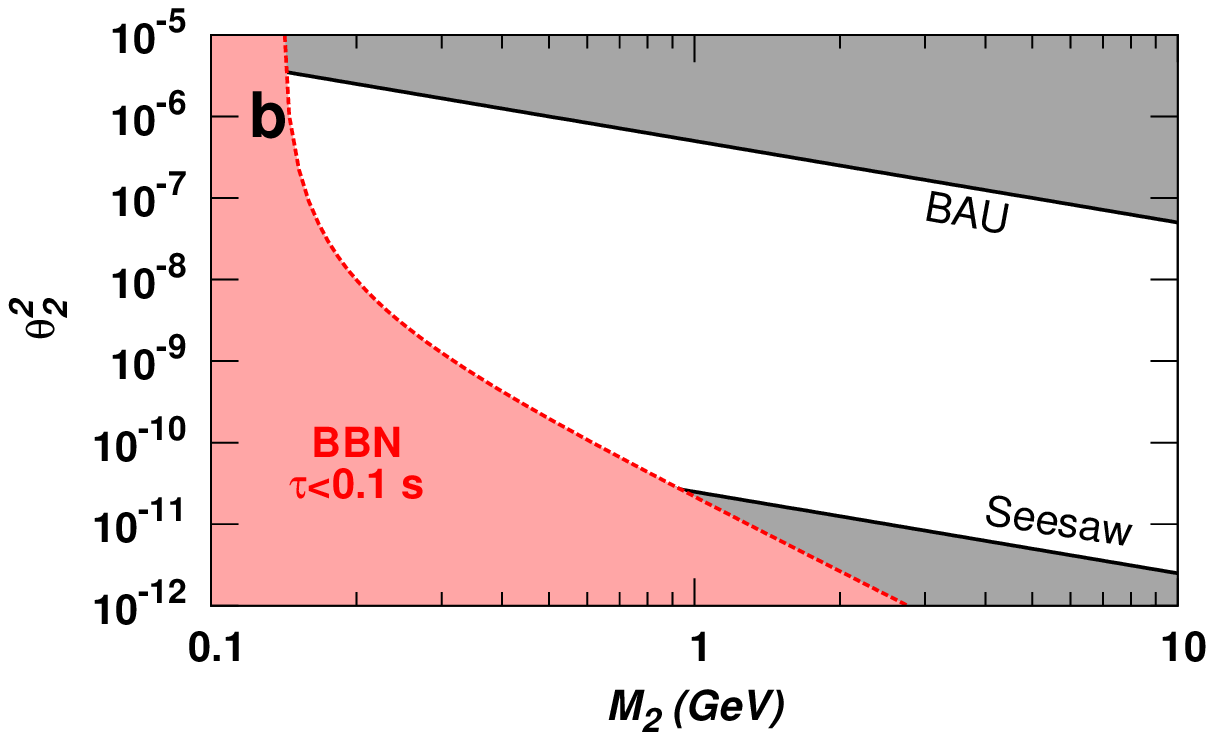}
  \caption{Constraints on the masses and mixing angles of the dark matter
    sterile neutrino $N_1$ (\textsl{a}) and of two heavier sterile neutrinos
    $N_{2,3}$ (\textsl{b}). These constraints come from astrophysics,
    cosmology, and neutrino oscillation experiments.  Abbreviations: BAU,
    baryon asymmetry of the Universe; BBN, big bang nucleosynthesis.}
\label{fig:sterile}
\end{figure}

The masses of these two sterile neutrinos should lie in the range $150\mev
\lesssim M_{2,3} \lesssim 100\gev$ and should be degenerate ($\Delta
M_{2,3}\ll M_{2,3}$).  These characteristics are required for baryon asymmetry
generation (Section~\ref{sec:bau}). In order not to affect the predictions of
big bang nucleosynthesis, we choose their masses and Yukawa couplings so that
their lifetime $\tau < 0.1$~s (i.e. they decay at temperatures $T\gtrsim
3$~MeV).  These bounds are summarized in Figure~\ref{fig:sterile}.

The presence of these sterile neutrinos can affect DM production.  If these
particles decay \emph{before} the DM is produced, they can 
generate significant lepton asymmetry. In the presence of lepton asymmetry
sterile neutrino production proceeds differently~\cite{Shi:98}. If $N_{2,3}$
decay \emph{after} DM production, they heat the primordial plasma, increasing
its entropy by a factor of $S$ and changing its temperature by $S^{1/3}>1$.
As the DM sterile neutrinos decouple from the primeval plasma, they take no
part in this entropy release, making their average momentum smaller than that
of the active neutrinos by the same factor, $S^{1/3}$. As a result, the
concentration of DM particles, compared to that of active neutrinos, decreases
by a factor of $S$,
and therefore the mixing angle, required to produce the same amount of DM
particles \emph{today}, becomes $S$ times bigger for the same DM mass
(compared to the $S=1$ case)~\cite{Asaka:06}. \emph{Therefore, the computation
  of sterile neutrino DM production and relation between the mass and mixing
  angle should be performed in the full \numsm}, taking into account the
influence of the neutrinos $N_{2},N_3$.

\section{The \numsm cosmology}
\label{sec:cosm-early-univ}

In this section we describe the history of the Universe in the \numsm from
inflation onward.

\subsection{Inflation in the \numsm}
\label{sec:inflation}
It has been shown~\cite{Bezrukov:07} that it is possible to incorporate
inflation within the \numsm approach, that is without introducing  new
particles above the electroweak scale. The role of the inflaton is played by
the SM Higgs field $\phi$. The non-minimal coupling of $\phi$ to the Ricci
scalar $R$
\begin{equation}
  \label{eq:21}
  \mathcal{L}_R = \xi \phi^\dagger\phi R\;,
\end{equation}
leads to a sufficiently flat effective potential for large values of $\phi$.
The constant $\xi$ is fixed by the Higgs mass and by the amplitude of scalar
fluctuations, obtained from COBE's observations of the cosmic microwave
background (CMB).  For the SM model to be a consistent field theory all the
way up to the Planck scale, the mass of the Higgs boson must lie in the
interval $126~{\rm GeV}< M_H < 194~{\rm
  GeV}$~\cite{Bezrukov:07,Bezrukov:08b,Bezrukov:09a} (see
also~\cite{DeSimone:2008ei,Barvinsky:2009fy}). For uncertainties in these
estimates see~\cite{Bezrukov:09a}. After inflation the Universe heats up to
the temperature $T = T_{\rm reh} > 1.5\times 10^{13}$~GeV, creating all of the
particles included in the SM.  However, the singlet fermions of the \numsm are
not created either during inflation or during reheating of the Universe
because of the small values of their Yukawa couplings~\cite{Bezrukov:08a}.

\subsection{Baryon asymmetry of the Universe}
\label{sec:bau}

Let us consider  the stage of the Universe's evolution between the reheating
temperature of $ \sim 10^{13}$ GeV and the temperatures of $\sim 100$~GeV. Our
Universe is baryon asymmetric -- it does not contain antimatter in amounts
comparable with matter, to wit:
\begin{equation}
  \label{eq:12}
  \eta_B\equiv \frac{n_B}{s} = (8.8\pm 0.2 ) \times 10^{-11}
\end{equation}
where $s$ is the entropy density~\cite{Dunkley:2008ie}.  To produce baryon
asymmetry of the Universe three conditions should be
satisfied~\cite{Sakharov:1967dj}:
\begin{inparaenum}[\em (a)]
\item Conservation of the baryon number should be violated,
\item CP symmetry should be broken, and
\item the corresponding processes should be out of thermal equilibrium.
\end{inparaenum}
Although the SM fulfills all there requirements~\cite{Kuzmin:1985mm}, it
cannot produce BAU because there is no first-order electroweak phase
transition with experimentally allowed Higgs boson masses
\cite{Kajantie:1996mn}. In addition, the CP violation in
Cabibbo--Kobayashi--Maskawa mixing of quarks is unlikely to lead to the
observed value of BAU~(\ref{eq:12}).

The only source of baryon number non-conservation in the \numsm (as in the SM)
is the electroweak anomaly~\cite{Kuzmin:1985mm}. The field configurations
important for baryon number non-conservation are known as
\emph{sphalerons}~\cite{Klinkhamer:1984di}.  The sum of baryon and
lepton numbers is not conserved at temperatures above $T_{\rm sph} \sim
100$~GeV~\cite{Kuzmin:1985mm}, and the BAU can be generated from
the lepton asymmetry at $T>T_\mathrm{sph}$.

A detailed description of the system of singlet leptons and active fermions in
the early Universe is necessarily quite complicated. The number of relevant
zero-temperature degrees of freedom (three active neutrinos, three sterile
neutrinos, and their antiparticles) is large, and the timescales of different
processes can vary by many orders of magnitude.  To understand the qualitative
picture of the creation of the matter excess (see
Refs.~\cite{Asaka:05b,Shaposhnikov:08a} for a detailed quantitative discussion
and~\cite{Akhmedov:98} for the original proposal of baryogenesis in
singlet-fermion oscillations), one must determine the reasons for departure
from thermal equilibrium \cite{Sakharov:1967dj}.  The singlet fermions, with
their very weak couplings to the SM fields, play an important role. The
production rate of $N_I$ is of the order of
\begin{equation}
\Gamma_I \sim |F_I|^2 T \;,
\end{equation} 
where $|F_I|^2 \equiv
\sum_\alpha |F_{\alpha I}|^2$. Because at $T_{\rm reh}$ singlet fermions were
absent, the number of created singlet fermions at temperature $T$ is of the
order of 
\begin{equation}
  q(T)\equiv\frac{n_I}{n_{eq}} \sim \frac{|F_I|^2 M_0}{T}~,
\label{eq}
\end{equation}
where $n_{eq}$ is an equilibrium concentration at given temperature. The
temperature-time relation is given by $t=\frac{M_0}{2T^2},~~ M_0\simeq
M_{Pl}/1.66\sqrt{g_{eff}}$, where $g_{eff}$ is the number of effectively
massless degrees of freedom. The relation shown in Eq.~(\ref{eq}) is valid
only if $q<1$.  The singlet fermion equilibrates at temperature $T_+$, defined
via $q(T_+)\sim 1$, as
\begin{equation}
  T_{+} \simeq |F_I|^2 M_0 \label{eq:17}.
\end{equation}

The phase-space density constraints, together with X-ray bounds
(Section~\ref{sec:dm}), place an upper bound on the Yukawa coupling of the
$N_1$ at $|F_1|^2 < 10^{-21}$.  The relation (\ref{eq}) shows that $N_1$ is
irrelevant for baryogenesis because $q_{N_1}\ll 1$ at all temperatures $T\ge
T_\mathrm{sph}$. In other words, $N_1$ practically decouples from the plasma.

If for both $N_2$ and $N_3$ $q(T_{\rm sph})>1$, $N_2$ and $N_3$ come to
thermal equilibrium before $T_{\rm sph}$.  In this case, even if baryon
asymmetry of the Universe were generated before equilibration, it would be
destroyed in equilibrium reactions with participation of sphalerons and
singlet fermions.  If $q(T_\mathrm{sph}) \sim 1$, the fermions approach
thermal equilibrium at the electroweak epoch, exactly what is needed for
effective baryogenesis.  This requirement restricts the masses of $N_{2,3}$ to
the $1-100$-GeV region. A detailed analysis of this condition, which accounts
for numerical factors and complicated flavor structure of asymmetry has been
performed~\cite{Shaposhnikov:08a}. It leads to a constraint
(Figure~\ref{fig:sterile}) on the mass-mixing angle plane.

Let us turn now to the effects of CP violation. Although the Yukawa matrix of
singlet fermions generally contains CP-violating phases, in the tree-level
processes their effects cancel out.  Any quantum loop, which is necessary for
CP violation to show up, is parametrically suppressed by
$F_{2\alpha}^2,F_{3\alpha}^2$, which are much smaller than $\eta_B$ (cf.
Eq.~(\ref{eq:10})). 
The only means of overcoming this factor is to have a resonance, which occurs
when the masses of sterile neutrinos $N_2$ and $N_3$ are nearly degenerate.

Let $\Delta M(T)$ be the mass difference between $N_2$ and $N_3$ (temperature
dependent due to Higgs' vev $T$ dependence and radiative corrections at high
temperatures).  The frequency of oscillations between $N_2$ and $N_3$ is given
by
\begin{equation} 
\label{eq:23}
\omega \sim \frac{|M_2^2 - M_3^2|}{E_I}\sim \frac{M_{2}\Delta M(T)}{T}\;,
\end{equation}
where we take into account that the typical energy of a singlet fermion
$E_I\sim T$ and that $\Delta M(T) \ll M_{2}\approx M_3$. If the rate of
oscillations exceeds the rate of the Universe's expansion $H(T)$, the system
is out of resonance, and no amplification of CP violation occurs. In the
opposite case, CP violation has no time to develop. This allows us to
determine the temperature $T_B$ at which baryogenesis occurs: $\omega \sim
H(T_B)$, which leads to~\cite{Akhmedov:98,Asaka:05b},
\begin{equation} 
  T_B \sim \Bigl(M_I \Delta M(T) M_0\Bigr)^{1/3}~.
\end{equation}
The described mechanism is quite effective. The \emph{maximal} baryon
asymmetry $\Delta_B \equiv \frac{n_B-n_{\bar B}}{n_B+n_{\bar B}}\sim 1$ (i.e.
$\eta_B \sim 10^{-2}$) is produced when $T_B \sim T_{\rm sph}\sim T_+$,
leading to $\Delta M(T_B) \sim 0.01\ev \left(\frac{1\gev}{M_I}\right)$.  It is
important that interactions with plasma do not destroy the quantum-mechanical
nature of oscillations of $N_{2,3}$, in other words that $T_{+}\lesssim
T_{B}$.

The overall qualitative picture is as follows: Coherent pairs of singlet
fermions are constantly created in interactions with the electroweak plasma.
These fermions oscillate in a CP violating manner.  As a result the Universe
develops lepton asymmetry in the active, left-handed sector. This asymmetry is
converted to baryon asymmetry due to anomalous electroweak reactions with
baryon number non-conservation.

Probably, the simplest way to deal with all these effects in a quantitative
way is to use the equation for the density matrix $\rho$
\cite{Dolgov:1980cq,Sigl:1992fn,Akhmedov:98,Asaka:05b}. 
In this case $\rho$ is a $12 \times 12$ matrix ($12=3\times 2\times 2$ degrees
of freedom for all active and sterile neutrino states). For the problem at
hand, this equation can be simplified (for details see Refs.~\cite{Asaka:05b,Shaposhnikov:08a}):

\begin{compactenum}[1.]
\item The rates of interactions between active neutrinos are much higher than
  the rate of the Universe's expansion. Therefore, coherent effects for active
  neutrinos are not essential, and the part of the general density matrix
  $\rho$ related to active leptonic flavors can be replaced with equilibrium
  concentrations characterized by three dimensionless chemical potentials
  $\mu_\alpha$ (the ordinary chemical potential divided by the temperature),
  giving the leptonic asymmetry in each flavor.
\item Active neutrinos acquire temperature dependent
  masses~\cite{Weldon:1982bn} that are much larger than
  those of singlet fermions.  Therefore, all non-diagonal elements of the
  density matrix simultaneously involving the active and sterile states can be
  set to zero.
\item The DM neutrino $N_1$ is completely decoupled from the system.
\end{compactenum}
This leaves us with the $2\times2$ density matrix $\rho_N$ for singlet
fermions $N_2$ and $N_3$, the charge conjugated density matrix $\bar{\rho}_N$
for corresponding antiparticles (opposite-chirality states, to be precise),
and three chemical potentials $\mu_\alpha$. The corresponding equations can be
written as \cite{Asaka:05b,Shaposhnikov:08a}:
\begin{eqnarray}
\label{kineq1}
i \frac{d\rho_N}{dt}&=& [\CH, \rho_N]
-\frac{i}{2}\{\Gamma_N, \rho_N - \rho^{eq}\} +
i \mu_\alpha{\tilde\Gamma^\alpha}_N~,\\
i \frac{d\bar\rho_N}{dt}&=& [\CH^*, \bar\rho_N]
-\frac{i}{2}\{\Gamma^*_N, \bar\rho_N - \rho^{eq}\} -i \mu_\alpha{
  \tilde\Gamma^{\alpha *}}_N~,\text{ and}
\label{kineq2}\\
i \frac{d\mu_\alpha}{dt}&=&-i\Gamma^\alpha_L\mu_\alpha +
i {\rm Tr}\left[{\tilde \Gamma^\alpha}_L(\rho_N -\rho^{eq})\right] -
i {\rm Tr}\left[{\tilde \Gamma^{\alpha*}}_L(\bar\rho_N -\rho^{eq})\right]
~.
\label{kineq3}
\end{eqnarray}
Here, $\CH=p(t)+\CH_0+\CH_{int}$ is the Hermitian effective Hamiltonian
incorporating the medium effects on neutrino propagation; $p(t)$ is the
neutrino momentum, with $\langle p(t)\rangle \sim 3T$ (we assume that all the
neutral fermion masses are much smaller than the temperature); $\CH_0 =
\frac{M^2}{2p(t)}$ (we include $\Delta M_{IJ}$ to $\CH_{int}$);
$\rho^{eq}=\exp{(-p/T)}$ is an equilibrium diagonal density matrix; and
$[~,~]$ ($\{~,~\}$) corresponds to the commutator (anti-commutator).  In
Eq.~(\ref{kineq3}) there is no summation over $\alpha$ and $\Gamma^\alpha_L$
are real. The explicit expressions for the equilibration rates
$\Gamma_N,~{\tilde\Gamma^\alpha}_N,~ \Gamma^\alpha_L,~{\tilde
  \Gamma^\alpha}_L$ via Yukawa couplings can be found in Ref.~\cite{Asaka:05b}
for the case in which the temperature is higher than the electroweak scale and
in Ref.~\cite{Shaposhnikov:08a} for the case in which the temperature is
smaller. 
These equilibration rates are all related to the absorptive parts of the two
point functions for active or sterile neutrino states, and they contain a
square of Yukawa couplings $F_{\alpha I}$.  The real parts of the
corresponding graphs, together with the mass-squared difference between $N_2$ and
$N_3$ determine the effective Hamiltonian $\CH$.  For high temperatures
$T\gsim T_{{\rm sph}}$ the equilibration processes are associated with Higgs,
$W$ and $Z$ decays to singlet and active fermions, 
to corresponding inverse
processes, and 
to $t\bar t\to N\bar\nu$ scattering (where $t$ is the top-quark). At smaller
temperatures $T\lsim T_{{\rm sph}}$ the rates are associated with $W$ and $Z$
exchange and with singlet-active mixing through the Higgs vev. The last terms
in Eqs.~(\ref{kineq1}) and (\ref{kineq2}), as well as Eq.~(\ref{kineq3}) are
crucial for the generation of lepton asymmetry
(see~\cite{Asaka:05b,Shaposhnikov:08a}).  In the earlier
work~\cite{Akhmedov:98} the computations and qualitative discussion were based
on incomplete kinetic equations, that lacked these terms.

Equations~(\ref{kineq1}--\ref{kineq3}), supplemented by initial conditions
$\rho_N = \bar\rho_N = \mu_\alpha = 0$ fixed by inflation
(Section~\ref{sec:inflation}), give a basis for the analysis of the baryon and
lepton asymmetry generation. It has been shown in
Refs.~\cite{Asaka:05b,Shaposhnikov:08a} that the observed baryon
asymmetry~(\ref{eq:12}) can be generated in a wide range of parameters of the
\numsm, provided $N_{2,3}$ are sufficiently degenerate:
\begin{equation}
  \label{eq:16}
  \Delta M(T_B) \sim \sqrt{|\Delta m_{\rm
      atm}^2|}\times\left(\frac{M_2}{\mathrm{GeV}}\right)^2 \times \left\{
    \begin{array}{ll}
      800, & \text{normal hierarchy}\\
      6400, & \text{inverse hierarchy}
    \end{array}\right.\;.
\end{equation}

The generation of baryon asymmetry stops at the temperature of the sphaleron
freeze-out $T_{\rm sph}$. The generation of the lepton asymmetry may still
occur at $T < T_{\rm sph}$~\cite{Shaposhnikov:08a}.  Moreover, the magnitude
of low-temperature lepton asymmetry has nothing to do with baryon asymmetry
and may exceed it by many orders of magnitude.  Sufficiently large lepton
asymmetry $\Delta_L \equiv
\frac{n_{\nu_e}-n_{\bar\nu_e}}{n_{\nu_e}+n_{\bar\nu_e}} > 10^{-3}$ can affect
the production of the DM sterile neutrino (see
Section~\ref{sec:dm-production-numsm}).\footnote{We denote by $n_{\nu_e}
  (n_{\bar\nu_e})$ the number density of the electron neutrino (antineutrino),
  and assume throughout the paper that the lepton asymmetry is flavor blind.}

Two different mechanisms  can lead to late lepton asymmetry
generation. As discussed above, the singlet fermions enter into
thermal equilibrium at $T_{+}$. As the temperature decreases below
the electroweak scale, the rate of interactions 
of singlet fermions with plasma gets suppressed by the Fermi constant, and
these fermions decouple at some temperature $T_-$, that is somewhat higher
than their mass~\cite{Shaposhnikov:08a}. The lepton asymmetry can then be
generated below $T_-$ in their oscillations, just as the baryon asymmetry was
created above the electroweak scale. Alternatively, the lepton asymmetry can
be created via decays of $N_{2,3}$. The analysis~\cite{Shaposhnikov:08a} shows
that large lepton asymmetries ($\Delta_L > 10^{-3}$) can be produced, provided
that the zero-temperature mass degeneracy of $N_{2,3}$ is high enough: $\Delta
M(0) \lesssim 10^{-4} \sqrt{|\Delta m^2_{\rm atm}|}$ (note, that $\Delta M(0)$
can differ significantly from $\Delta M(T_B)$.) The \emph{maximal possible
  lepton asymmetry} that can be produced at $T_-$ or during the decays of
$N_{2,3}$ is $\Delta_L^\text{max}=4/(9\times2+4)=2/11$, where $4$ is the total
number of spin-states of $N_{2,3}$ and $9$ is the number of spin-states of
three lepton generations. This degree of asymmetry can be attained if
CP violation is maximal.  Generation of the baryon asymmetry at the
level shown in Eq.~(\ref{eq:12}), together with the large lepton asymmetry, is possible for
the singlet fermion masses in the ${\cal O}({\rm GeV})$ range and $|F_2|\sim
|F_3|$.  Following the Refs.~\cite{Shaposhnikov:08a,Laine:08a} we  characterize the
lepton asymmetry as $L_6\equiv 10^6 (n_{\nu_e}-n_{\bar\nu_e})/s$. 
The value $\Delta_L^\text{max}=2/11$ corresponds to $L_6^\text{max} \approx
700$.  This value is still considerably smaller than the upper limit, imposed
by the big bang nucleosynthesis~\cite{Serpico:05} $L_6^\textsc{bbn}\approx
2500$.

\subsection{DM production in the \numsm}
\label{sec:dm-production-numsm}

In the \numsm  DM sterile neutrinos are produced in the early Universe due
to their coupling to active neutrinos.  An estimate of the rate of DM sterile
neutrino production $\Gamma_N$ at temperatures below the electroweak scale is
given by \cite{Barbieri:1989ti}
\begin{equation}
\Gamma_N \sim \Gamma_\nu \theta_M^2(T)~,
\label{st}
\end{equation} 
where $\Gamma_\nu\sim G_F^2 T^5$ is the rate of active neutrino production,
and where $\theta_M(T)$ is a temperature- (and momentum-) dependent mixing angle:
\begin{equation}
  \theta_1^2 \rightarrow \theta_M^2(T)
  \simeq \frac{\theta_1^2}{\Bigl(1+ \frac{2 p}{M_1^2}\bigl(
    b(p,T)\pm c(T)\bigr)\Bigr)^2+\theta_1^2}~.
\label{effang}
\end{equation} 
Here~\cite{Notzold:87}
\begin{equation}
  b(p,T)=\frac{16
    G_F^2}{\pi\alpha_W}p(2+\cos^2\theta_W)\frac{7\pi^2T^4}{360},~~
  c(T)=3\sqrt{2}G_F \Bigl(1+\sin^2\theta_W\Bigr)(n_{\nu_e}-n_{\bar\nu_e})~, 
\end{equation}
where $\theta_W$ is the weak mixing angle, $\alpha_W$ is the weak coupling
constant, and $p\sim \text{few}\;T$ is the typical momentum of created DM
neutrinos. The term $c(T)$ in~(\ref{effang}) is proportional to the
\emph{lepton asymmetry} and contributes 
with the opposite sign to the mixing
of $N_1$ with active neutrinos and antineutrinos.

If the term $b(p,T)$ dominates $c(T)$ for $p\sim (2-3)T$ [which we  refer to
as non-resonant production \emph{NRP}), the production
rate~(\ref{st}) is strongly suppressed at temperatures above a few hundred
MeV and peaks roughly at \cite{Dodelson:93}
\begin{equation}
  T_{peak} \sim 130\left(\frac{M_1}{1\kev}\right)^{1/3}~\mbox{MeV}~,
  \label{peak}
\end{equation}
corresponding to the temperature of the quantum chromodynamics crossover for
keV scale sterile neutrinos.  This complicates the analysis, because neither
the quark, nor the hadron description of plasma is applicable at these
temperatures~\cite{Asaka:06b}.  In the region of the parameter space
$(\theta_1,~M_1)$, as demonstrated by X-ray observations together with dSph
constraints (Section~\ref{sec:dm}) $\Gamma_N(T_{peak}) \ll H(T_{peak})$. In
other words the DM sterile neutrinos were never in thermal equilibrium in the
early Universe.  This fact allows us to make first principle computation of
their abundance via certain equilibrium finite temperature Green's functions
computed in the SM without inclusion of right-handed
fermions~\cite{Asaka:06b,Asaka:06c,Laine:08a}.  (See
Refs.~\cite{Abazajian:01a,Abazajian:02,Abazajian:05a} for earlier computations
of sterile neutrino DM abundance. The sterile neutrino oscillations in the
medium were also considered in Ref.~\cite{Boyanovsky:07a}).

The shape of the primordial momentum distribution of the NRP sterile neutrinos
is roughly proportional to that of the active neutrinos~\cite{Dolgov:00}:
\begin{equation}
  \label{eq:19}
  f_{N_1}(t,p) = \frac{\chi}{e^{p/T_\nu(t)}+1}\;,
\end{equation}
where the normalization $\chi \sim \theta_1^2$ and where $T_\nu(t)$ is the
temperature of the active neutrinos. The true distribution of sterile
neutrinos is in fact colder than that shown in Eq.~(\ref{eq:19}).
Specifically, the maximum of $p^2 f_{N_1}(p)$ occurs at $p/T_\nu\approx
1.5-1.8$ (depending on $M_1$), as compared with $p\approx 2.2 T_\nu$ for the
case shown in Eq.~(\ref{eq:19}) \cite{Asaka:06b,Asaka:06c}.

The NRP mechanism specifies a \emph{minimal} amount of DM that will be
produced for given $M1$ and $\theta_1$. The requirement that 100\% of DM be
produced via NRP places an \emph{upper bound} on the mixing angle $\theta_1$
for a given mass. This conclusion can only be affected by entropy dilution
arising from the decay of some heavy particles below the temperatures given in
Eq.~(\ref{peak})~\cite{Asaka:06}. In the \numsm such dilution may occur due to
the late decay of two heavier sterile neutrinos (see
Section~\ref{sec:two-heavy-nu}).

The production of sterile neutrino DM may substantially change in the presence
of lepton asymmetry. If the denominator in Eq.~(\ref{effang}) is small,
\begin{equation}
  1+ 2p\frac{b(p,T)\pm c(T)}{M_1^2}=0,\label{eq:15}
\end{equation}
then resonant production (\emph{RP}) of sterile neutrinos \cite{Shi:98}
occurs, analogous to the Mikheyev--Smirnov--Wolfenstein
effect~\cite{Wolfenstein:1977ue,Mikheev:1986gs}. In this case the dispersion
relations for active and sterile neutrinos cross each other at some momentum
$p$, determined from Eq.~(\ref{eq:15}). This results in the effective transfer
of an excess of active neutrinos (or antineutrinos) to the population of DM
sterile neutrinos.  The maximal amount of sterile neutrino DM that can be
produced in such a way ($n_{res}$) is limited by the value of lepton
asymmetry: $n_{res} \lesssim |n_{\nu_e}-n_{\bar\nu_e}|$. The present DM
abundance $\Omega_\dm \sim 0.2$~\cite{Dunkley:2008ie} translates into
$n_{res}/s \lesssim 10^{-6}\frac{\mathrm{keV}}{M_1}$, in other words \emph{the resonant
  effects in production of sterile neutrinos are not essential for} $L_6
\lesssim 1$ (for masses in the keV range).

\begin{figure*}
  \begin{tabular}{lc}
    \begin{minipage}{0.49\linewidth}
      \hskip -4em \includegraphics[width=1.45\linewidth]{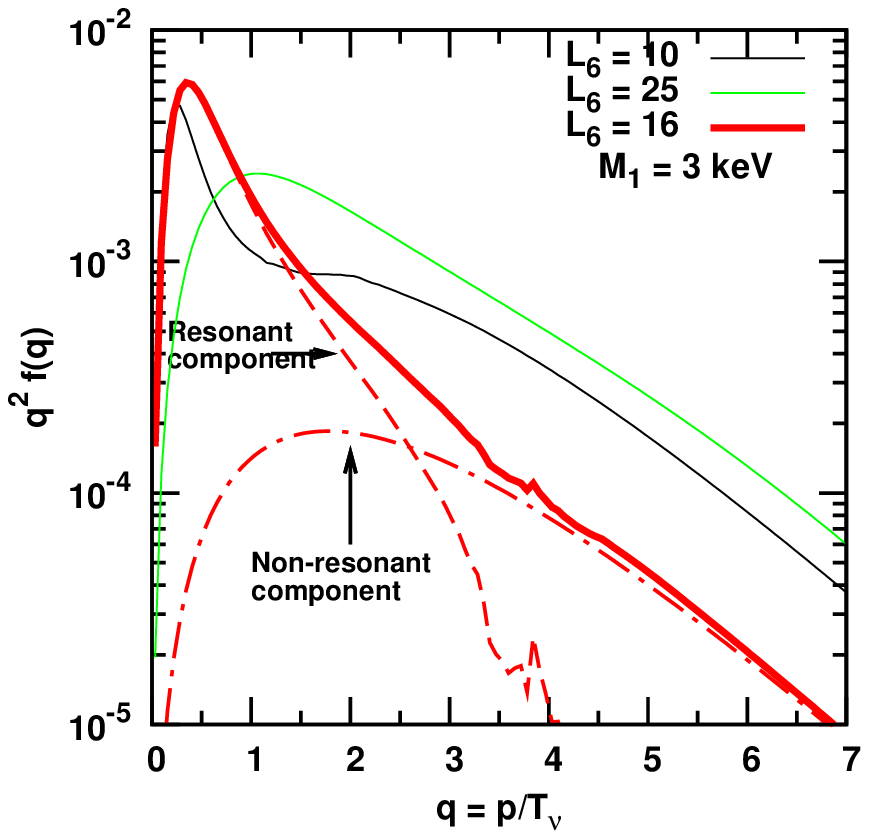}
    \end{minipage}
    &
    \begin{minipage}{0.49\linewidth}
      \includegraphics[width=\linewidth]{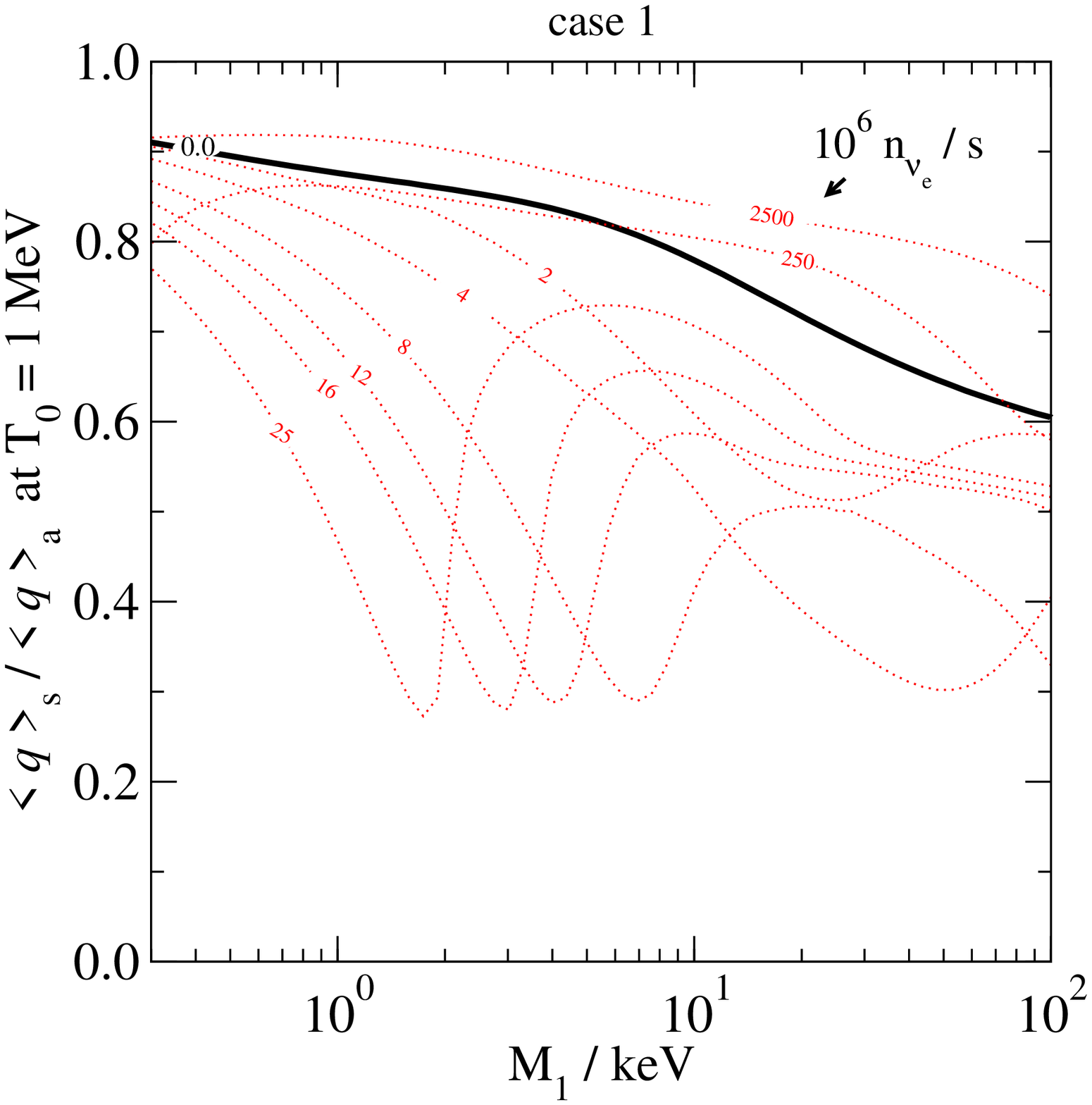} %
    \end{minipage}\\
    \multicolumn{1}{c}{(a)} & (b)\\
  \end{tabular}
  \caption{\textbf{(a)} Characteristic form of the resonantly produce (RP)
    sterile neutrino spectra for $M_1 =3\kev$ and $L_6=16$ (\textsl{red solid
      line}) and its resonant and nonresonant components. Also shown are
    spectra for $L_6=10$ and $L_6=25 $, all of which have the same form for
    $q\gtrsim 3$. All spectra are normalized to give the same DM abundance.
    For the spectrum with $L_6=16$, its nonresonantly produce (NRP) component
    $\fnrp \simeq 0.12$ (\textsl{(red dashed-dotted line}) and its colder, RP
    component (\textsl{red dashed line}) are shown separately.  The spectrum
    with $L_6=10$ has $\fnrp \simeq 0.53$ and $L_6=25$ has $\fnrp \simeq
    0.60$.  The RP components happen to peak around $q_\text{res}\sim 0.25-1$.
    Their width and height depends on $L_6$ and $M_1$.  \textbf{(b)}
    Dependence of the ratio of the average momentum of sterile neutrinos to
    that of the active neutrinos on mass $M_1$ and lepton asymmetry. The black
    solid curve represents the ratio of average momenta, computed over the
    exact NRP spectra to the ratio using the approximate analytic form shown
    in Eq.~(\ref{eq:19}).  The black solid curve is the ratio of average
    momenta, computed over exact NRP spectra to the ratio, computed using the
    approximate analytic form~(\ref{eq:19}).  For each mass there is lepton
    asymmetry for which the $\langle q\rangle$ reaches its minimum $\langle
    q\rangle_{min} \sim 0.3\langle q\rangle_{\nu_\alpha}$.  The corresponding
    spectra are the most distinct from the NRP spectra. The fraction $\fnrp$
    for the former can be $\lesssim 20\%$.  Panel \textbf{b} reproduced from
    Ref.~\cite{Laine:08a} with permission.}  \label{fig:rp-spectra}
\end{figure*}

Resonantly produced DM sterile neutrinos have spectra that significantly
differ from those in the NRP case~\cite{Shi:98,Laine:08a}. As a rule, the
primordial velocity distribution of RP sterile neutrinos contains narrow
resonant (\emph{cold}) and nonresonant (\emph{warm}) components.  The spectra
for $M_1 = 3\kev$ for $L_6=10,16,25$ are shown in Fig.~\ref{fig:rp-spectra}a
as functions of comoving momentum $q=p/T_\nu$. The non-resonant tails are
self-similar for the same mass and for different lepton asymmetries, and for
$q\gtrsim 3$ they behave as rescaled NRP spectra~\cite{Asaka:06c} with the
same mass.  We term this rescaling coefficient \emph{the fraction of the NRP
  component} $\fnrp$.  The dependence of the average momentum $\langle
q\rangle$
on the
mass and lepton asymmetry is shown in Fig.~\ref{fig:rp-spectra}b.

 \subsection{Big Bang Nucleosynthesis.}
\label{sec:constraints-from-big} Decays of heavy singlet fermions $N_{2,3}$
heat the primeval plasma and distorts the spectra of active neutrinos from
their thermal values. If this distortion occurs around the time of BBN, it may
significantly influence its predictions. It has been
shown~\cite{Dolgov:00a,Dolgov:00b} that if the lifetime of sterile neutrinos
$\tau_{2,3} < 0.1$~s (i.e.  they decay at temperatures above $\sim 3$~MeV),
the decay products have enough time to thermalize without spoiling the
nucleosynthesis.  This requirement places a \emph{lower} bound on the strength
of Yukawa interactions $F_{\alpha 2}$ and $F_{\alpha 3}$
(Fig.~\ref{fig:sterile})~\cite{Gorbunov:07a}. A detailed treatment of the
decays of sterile neutrinos and their interaction with primordial plasma can
somewhat change these constraints, in particular by allowing for smaller
Yukawa couplings.  \subsection{Structure formation in the \numsm}
\label{sec:structure-form} 
\begin{figure}
  \centering\includegraphics[width=\linewidth]{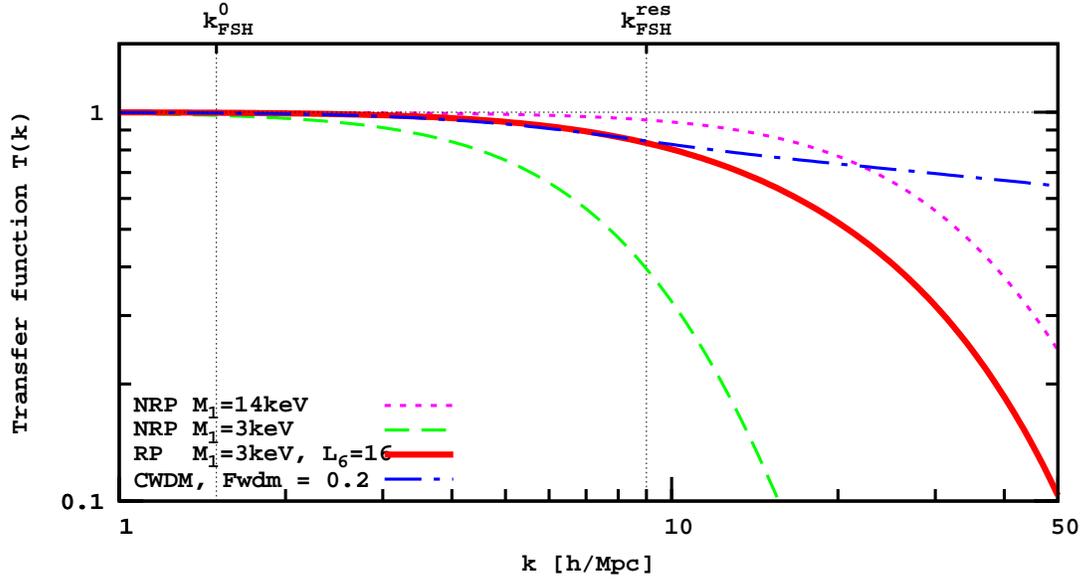} %
  \caption{Transfer functions (TFs) for the nonresonant production (NRP)
    spectra where $M_1=14\kev$ and $M_1=3\kev$. Also shown is a resonant
    production (RP) spectrum where $M_1=3\kev$ and $L_6=16$ together with a
    cold/warm DM spectrum for $M = 3\kev$ and $F_\wdm \simeq 0.2$
    (\textsl{blue dashed-dotted line}). Vertical lines mark the free-streaming
    horizon of the nonresonant ($k^0_\textsc{fsh}$, left line) and resonant
    ($k_\textsc{fsh}^\text{res}$, right line) components. Abbreviations: CWDM,
    cold-plus-warm dark matter; WDM, warm dark matter.}  \label{fig:rp-tf}
\end{figure}

Comparing the production temperatures
Eq.~(\ref{peak}) of DM sterile neutrinos with their masses shows that they are
produced relativistically in the radiation-dominated epoch.  Relativistic
particles stream out of the overdense regions and erase primordial density
fluctuations at scales below the \emph{free-streaming horizon} (FSH) --
particles' horizon when they becomes nonrelativistic. (For a detailed
discussion of characteristic scales see e.g.~\cite{Boyarsky:08c} and
references therein). This effect influences the formation of structures. If DM
particles decouple nonrelativistically (\emph{cold} DM models, CDM) the
structure formation occurs in a ``bottom-up'' manner: specifically, smaller
scale objects form first and then merge into the larger
ones~\cite{Peebles:80}. CDM models fit modern cosmological data well.  In the
case of particles, produced relativistically and \emph{remaining relativistic}
into the matter-dominated epoch (i.e. \emph{hot} DM, HDM), the structure
formation goes in a ``top-down'' fashion~\cite{Zeldovich:70}, where the first
structures to collapse have sizes comparable to the Hubble
size~\cite{Bisnovatyi:80,Bond:80,Doroshkevich:81}. The HDM scenarios
contradict large-scale structure (LSS) observations.  Sterile neutrino DM are
produced out of equilibrium at the temperatures shown in Eq.~(\ref{peak}) and
is then redshifted to nonrelativistic velocities in the radiation-dominated
epoch (\emph{warm} DM, WDM). Structure formation in WDM models is similar to
that in CDM models at distances above the free streaming scale. Below this
scale density fluctuations are suppressed, compared with the CDM case. The
free-streaming scale can be estimated as~\cite{Bond:80} \begin{equation}
  \label{eq:5} \lambda_\textsc{fs}^{co} \sim 1\mpc
  \left(\frac{\mathrm{keV}}{M_1}\right)\frac{\langle p_N\rangle}{\langle
    p_\nu\rangle}\;.  \end{equation} where $1\mpc$ is the (comoving) horizon
at the time when momentum of active neutrinos $\langle p_\nu \rangle \sim
1\kev$. If the spectrum of sterile neutrinos is nonthermal, then the
non-relativistic transition and $\lambda_\textsc{fs}^{co}$ shift by $\langle
p_{N}\rangle/\langle p_\nu\rangle$.  To account for the influence of
primordial velocities on the evolution of density perturbations, it is
convenient to introduce a \emph{transfer function} (TF) with respect to the
CDM model, \begin{equation} T(k) \equiv
  \bigl[P_\wdm(k)/P_{\cdm}(k)\bigr]^{1/2}\le 1\label{eq:20}~.  \end{equation}
where the power spectra $P_\wdm(k)$ and $P_{\cdm}(k)$ are computed with the
same cosmological parameters. In general, the characteristic shape of a TF
depends upon a particular WDM model. In the \numsm the TF's shape is
determined by the mass of DM particle and lepton asymmetry, which define the
primordial velocity distribution.  In the case of both RP and NRP sterile
neutrino, the TF starts to deviated from one for $k\gtrsim k_\textsc{fsh}^0$,
where \begin{equation} \label{eq:13} k_\textsc{fsh}^0 = 0.5
  \left(\frac{M_1}{1~\mathrm{keV}}\right) \left(\frac{0.7}{h}\right) \left( 1
    + 0.085 \ln \left[ \left(\frac{0.1}{\Omega_M h^2}\right)
      \left(\frac{M_1}{1~\mathrm{keV}}\right) \right] \right)^{-1}
  h/\mathrm{Mpc}~.  \end{equation} is a \emph{free-streaming horizon} for a
model with mass $M_1$ and the velocity distribution in Eq.~(\ref{eq:19})
(c.f.~\cite{Boyarsky:08c}). In Eq.~(\ref{eq:13}) $\Omega_M$ is the total
matter density, and $H_0 =h \times 100\;\mathrm{km\:s^{-1}\:Mpc^{-1}}$ is the
Hubble constant today. For $k\gg k_\textsc{fsh}^0$ the TF of NRP sterile
neutrinos falls off as $\sim k^{-10}$
(c.f.~\cite{Bode:00,Viel:05,Boyarsky:08c}).  For the TF of RP sterile
neutrinos we define $k_\textsc{fsh}^\text{res} \simeq k^0_\textsc{fsh} \langle
q\rangle/q_\text{res}$. For scales $k\le k_\textsc{fsh}^\text{res}$ the
$T_\rp(k)$ is indistinguishable from that of cold-plus-warm dark matter (the
mixture of the WDM component in the form of NRP sterile neutrinos and pure
CDM) with the same mass and the same fraction of the warm component $F_\wdm =
\fnrp$. For smaller scales ($k> k_\textsc{fsh}^\text{res}$) the $T_\rp(k)$
decays, because the cold component of RP sterile neutrinos also has a
nonnegligible free-streaming, determined roughly by
$k_\textsc{fsh}^\text{res}$.  This characteristic behavior of NRP and RP
spectra is demonstrated in Fig.~\ref{fig:rp-tf}. We discuss cosmological
restrictions on the sterile neutrino DM models in
Section~\ref{sec:lya-constraints}.   \section{Search for \numsm physics} \label{sec:testing-numsm}
\subsection{Combined Restrictions from Astrophysics} \label{sec:restr-astro}
Below we summarize the astrophysical bounds on the sterile neutrino dark
matter.  \subsubsection{Phase-space density bounds.}  The most robust
\emph{lower bound} on the DM mass comes from the analysis of the phase-space
density of compact objects such as dwarf spheroidal galaxies of the Milky Way.
The universal Tremaine-Gunn bound~\cite{Tremaine:79} can be strengthened if
one \begin{inparaenum}[\bf (a)] \item knows the primordial velocity
  distribution of the DM particles and \item takes into account that the
  dynamics of DM particles is collisionless and dissipationless and that an
  initial phase-space volume occupied by DM particles is therefore preserved
  in the course of evolution. (For further discussion, see,
  e.g.,~\cite{Boyarsky:08a,Gorbunov:08b,Boyanovsky:08} and references
  therein).  \end{inparaenum} For the NRP sterile neutrino, this leads to $M_1
> 1.8\kev $~\cite{Boyarsky:08a}.  For the same mass, the primordial velocity
distribution of RP sterile neutrino DM is colder than the NRP dark matter.
Analysis of available spectra shows that models in which $M_1 > 1 \kev$ are
allowed for most values of lepton asymmetries~\cite{Boyarsky:08a}.  Note,
however, that in derivation of these bounds (unlike the bound, based on the
Pauli exclusion principle) the presence of dissipative baryons was not taken
into account.  However, because dSphs are the most dark and DM-dominated
objects (with a mass-to-light ratio $>100 M_\odot/L_\odot$), we expect the
results of Ref.~\cite{Boyarsky:08a} to be robust.  \subsubsection{X-ray
  constraints.}  \label{sec:x-ray-constraints} Because the sterile neutrino is
a decaying DM candidate, its decays would produce a narrow line in spectra of
DM-dominated astrophysical objects~\cite{Dolgov:00,Abazajian:01b}. The width
of this line is determined by the virial motion of DM particles in halos. It
ranges from $\Delta E/E \sim 10^{-2}$ for galaxy clusters to $10^{-4}$ for
dSphs. The expected flux of the DM decay is given by \begin{equation}
  \label{eq:11} F_\dm = \frac{M_\dm^\text{fov} \Gamma}{4\pi
    D_L^2}\frac{E_\gamma}{M_1} \simeq 6.38
  \left(\frac{M_{\dm}^\text{fov}}{10^{10}M_\odot}\right)
  \left(\frac{\!\mpc}{D_L}\right)^2 \times \sin^2(2\theta_1)
  \left[\frac{M_1}{\mathrm{keV}}\right]^5 \frac{\mathrm{keV}}{\mathrm{cm^2
      \cdot sec}}\;, \end{equation} where $M_\dm^\text{fov}$ is the mass of DM
within a telescope's field of view (FoV), the decay rate $\Gamma$ is given by
Eq.(\ref{eq:11}) and $D_L$ is the luminous distance to the object.
Eq.~(\ref{eq:11}) is valid in nearly all interesting cases apart from that of
the signal from the Milky Way halo (c.f.~\cite{Boyarsky:06c,Boyarsky:06d}).
We begin our discussion of X-ray constraints with considering the mass range
$0.4\kev \lesssim M_1 \lesssim 30\kev$ with corresponding photon energy
$E_\gamma = M_1/2$ falling into the X-ray band.  Modern X-ray instruments have
a FoV of $\sim 10'-15'$ which is below the characteristic scale of DM
distribution in a large variety of objects. As a result, the DM density does
not change significantly within the instrument's FoV, and the DM
flux~(\ref{eq:11}) is determined by the \emph{DM column density}
$\mathcal{S}=\int \rho_\dm(r) dr$: \begin{equation} \label{eq:8} F_\dm =
  \frac{\Gamma\Omega_\text{fov}}{8\pi}\hskip -3ex\int\limits_{\text{line of
      sight}} \hskip -3ex\rho_\dm(r)dr \;.  \end{equation} Note, that in this
limit the expected signal does not depend on the distance to the object.
Instead, it is determined by the angular size $\Omega_\text{fov}$ and the DM
overdensity $\mathcal{S}$.  The central DM column density is comparable
(within a factor $\sim 10$) for most of DM-dominated objects of different
types and scales. The objects with the highest column densities are central
regions of dSphs (such as Ursa Minor or Draco) and of Andromeda galaxy.  The
expected signals from these objects are a few times stronger than that of the
Milky Way halo (through which one observes all the objects). However there are
a variety of objects, covering the FoV of an X-ray instrument, that would
produce comparable signals (in contrast with signals from annihilating DM).
This gives us a certain freedom to choose observational targets, that may be
exploited in two ways (described below).  Galaxies and galaxy clusters are
X-ray bright objects whose spectra contain many atomic lines.  The DM decay
line should be distinguished from these lines (and from those of the
instrumental origin).  One can reliably identify atomic lines in cases where
several of them originate from transitions in the same element and therefore
have known relative intensities.  The emission of hot gas (including the
positions and intensities of lines) can be described by models with a few
parameters such as abundances of elements.  Another way to distinguish an
unidentified astrophysical line from the DM decay line is to study its
surface-brightness profile.  The surface brightness profiles of atomic lines
are proportional to the gas density squared, while the surface brightness
profile of a DM line is proportional to the density.  To avoid complicated
modeling and related uncertainties, one can analyze dark outskirts of extended
objects (such as clusters~\cite{Boyarsky:06b} or big
galaxies~\cite{Boyarsky:07a}) or search for a line against a featureless
astrophysical background such as the extragalactic X-ray
background~\cite{Boyarsky:05,Boyarsky:06c,Boyarsky:06d,Abazajian:06b}.  The
best strategy is to look in the spectra of X-ray quiet and DM dominated
objects, in particular dSphs~\cite{Boyarsky:06c,Boyarsky:06d}.  Extensive
searches for these lines have been performed using \emph{XMM-Newton}
\cite{Boyarsky:05,Boyarsky:06b,Boyarsky:06c,Watson:06,Boyarsky:06d,Boyarsky:06f,Boyarsky:07a},
\emph{Chandra}~\cite{Riemer:06,Watson:06,Boyarsky:06e,Abazajian:06b},
\emph{INTEGRAL}~\cite{Yuksel:07,Boyarsky:07b}, and
\emph{Suzaku}~\cite{Loewenstein:08}.  They did not reveal any ``candidate''
line in the energy range from $\sim 0.5\kev$ to about $14\mev$. These searches
yielded an upper bound on the possible DM flux.  To convert it to the upper
bound on the mixing angle (via Eq.~(\ref{eq:4})), one must reliably determine
the amount of DM in the observed objects.  Determining the amount of DM in any
given object is subject to specific systematic uncertainties, related e.g. to
assumptions about the spherical symmetry of the DM halo, modeling of the
contribution of baryons to the rotation curves, and assumptions about the
hydrostatic equilibrium of intra-cluster gas, etc (see
e.g.~\cite{Boyarsky:06e,Boyarsky:07a,Boyarsky:07b}).  Therefore, it is
important to rule out the same parameter values from several targets
(preferably of different scale and nature).

\begin{figure}
  \centering
  \includegraphics[width=\linewidth]{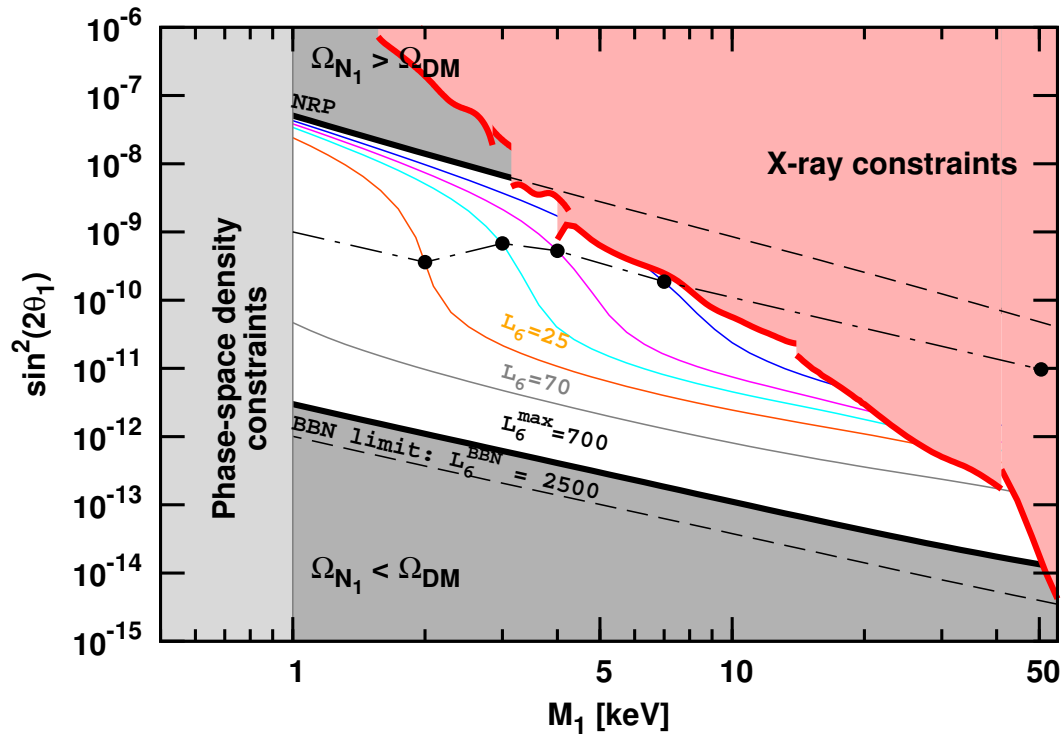}
  \caption{The allowed region of parameters for DM sterile neutrinos produced
    via mixing with the active neutrinos (\textsl{unshaded region}). The two
    thick black lines bounding this region represent production curves for
    nonresonant production (NRP) (\textsl{upper line}, $L_6=0$) and for
    resonant production (RP) (\textsl{lower line}, $L_6^{max}=700$) with the
    maximal lepton asymmetry, attainable in the
    \numsm~\cite{Laine:08a,Shaposhnikov:08a}.  The thin colored curves between
    these lines represent production curves for (\textsl{from top to bottom})
    $L_6 = 8,12,16,25,$ and $70$. The red shaded region in the upper right
    corner represents X-ray
    constraints~\cite{Boyarsky:06c,Boyarsky:06d,Boyarsky:07a,%
      Boyarsky:07b,Loewenstein:08} (rescaled by a factor of two to account for
    possible systematic uncertainties in the determination of DM
    content~\cite{Boyarsky:06e,Boyarsky:07a}).  The black dashed-dotted line
    approximately shows the RP models with minimal $\langle q\rangle$ for each
    mass, i.e., the family of models with the largest cold component.  The
    black filled circles along this line are compatible with the \lya
    bounds~\cite{Boyarsky:08d}, and the points with $M_1 \leq 4\kev$ are also
    compatible with X-ray bounds.  The region below $1\kev$ is ruled out
    according to the phase-space density arguments~\cite{Boyarsky:08a}.
    Abbreviation: BBN, big bang nucleosynthesis.}
  \label{fig:sf-window}
\end{figure}

The bounds obtained from the combined results of these searches
(Figure~\ref{fig:sf-window}) are based on independent results from various
objects (extragalactic X-ray background~\cite{Boyarsky:05,Boyarsky:06c}, the
Milky
Way~\cite{Boyarsky:06c,Riemer:06,Boyarsky:06d,Boyarsky:06f,Abazajian:06b,Yuksel:07,Boyarsky:07b},
M31~\cite{Watson:06,Boyarsky:07a}, galaxy
clusters~\cite{Boyarsky:06b,Boyarsky:06e}, and dSph
galaxies~\cite{Boyarsky:06d,Loewenstein:08}).  Superposing several results
makes them robust.  Comparing the of production curves (i.e. the relations
between $\theta_1$ and $M_1$, which produce the required DM abundance) with
the bounds on DM decay flux leads to the upper mass bound.  In the NRP case,
the \emph{upper bound} on the DM sterile neutrino mass was found to be $M_1
\le 4~ \kev$~\cite{Boyarsky:07a}. For a more effective RP scenario, the
smaller mixing angles are admissible, and the corresponding mass bound is much
weaker: $M_1 \lesssim 50$~keV~\cite{Laine:08a,Boyarsky:08a}.

To improve the modern astrophysical bounds (Fig.~\ref{fig:sf-window}) with
existing instruments, one should conduct prolonged (several Msec) observations
of dSph galaxies and other X-ray quiet objects, then combine various
observations. The bounds would improve roughly as $T_\text{exp}^{1/2}$ with
cumulative exposure time.

A real improvement in sensitivity can be achieved with the next generation of
X-ray spectrometers which will have an energy resolution $\Delta E/E \sim
10^{-3}\text{--}10^{-4}$ and which will maximize the product of their
effective area and FoV. For a comparison of the sensitivities of existing and
future missions, see Ref.~\cite{Boyarsky:06f,denHerder:09}.  For an example of
a spectrometer with optimal characteristics for performing a decaying DM search see
e.g.~\cite{Piro:08,denHerder:09}. Such a mission would allow us to improve the
sensitivity towards the DM search by up to a factor of $10^3$.  Its
preferred observational targets would be our Galaxy, its dwarf satellites, and
galaxy clusters.

If a candidate line is found, its surface-brightness profile can be
measured, distinguished from astrophysical lines, 
and compared with those of several objects with the same
expected signal.  This would make an astrophysical search for decaying DM
\emph{another type of a direct detection experiment}.

\subsubsection{\lya constraints.}
\label{sec:lya-constraints}

Below we describe the Lyman-alpha method and the Lyman-alpha constraints on the
properties of warm and cold-plus-warm dark matter models and in particular sterile
neutrino dark matter.

\paragraph{Structure formation at different scales.} The sterile neutrino DM
creates a cut-off in the power spectrum of matter density fluctuations at
sub-Mpc scales.  At larger scales, the power spectrum can be determined by CMB
experiments, or reconstructed from observed galaxy power spectrum 
(via LSS surveys). For any mass and mixing angle that satisfies the
constraints named above, the \emph{sterile neutrino DM model fits the CMB and
  LSS data} as well as the $\Lambda$CDM ``concordance model''.  One of the
main probes of the matter power spectrum at sub-Mpc scales is the
\emph{Lyman-$\alpha$ forest method}, discussed below.

\paragraph{\lya method.} It has been well established by analytical
calculations and hydrodynamical simulations that \lya absorption lines in
spectra of distant quasars are produced by clouds of neutral hydrogen at
different redshifts along the line of sight.  This neutral hydrogen is part of
the warm ($\sim 10^4$-K) and photoionized intergalactic medium. Opacity
fluctuations in the spectra arise from fluctuations in the neutral hydrogen
density, which can be used to trace cosmological fluctuations on scales $k
\sim (0.1-5) \, h \mpc^{-1}$, at redshifts $z \sim 2-4$. Thus it is possible
to infer fluctuations in the total matter
distribution~\cite{Bi:93,Viel:2001hd,Zaldarriaga:2001xs}. For each quasar, the
observed spectrum $I(z)$ can be expanded in (one-dimensional) Fourier space.
The expectation value of the squared Fourier spectrum is known as the \emph{flux
  power spectrum} $P_F(k)$.

The flux power spectrum $P_F(k)$ is a complicated function of the cosmological
parameters.  In the range of scales probed by \lya data, the effects of
free-streaming may be compensated for by a change in other cosmological
parameters. Therefore, it is important to perform collective fits to the \lya
data and to CMB and LSS data. At redshifts and scales probed by \lya, the
fluctuations already enter into a mildly non-linear stage of gravitational
collapse ($\delta \rho/ \rho\gtrsim1$).  To predict $P_F(k)$ for a given
cosmological model, one must perform hydrodynamical simulations: for CMB
experiments probing the linear matter power spectrum, it is sufficient to
compute the evolution of linear cosmological perturbations using a Boltzmann
code such as {\sc camb}~\cite{Lewis:99}).  Hydrodynamical simulations are
necessary both for simulating the non-linear stage of structure formation and
for computing the evolution of thermodynamical quantities (so as relate the
non-linear matter power spectrum to the observable flux power spectrum).  In
principle, to fit Lyman-$\alpha$ data one should perform a full
hydrodynamical simulation for each cosmological model. This is computationally
prohibitive, and various simplifying approximations have been
proposed~\cite{Theuns:98,Gnedin:01,McDonald:05,Viel:04,Viel:05,Viel:05b,Viel:05c,Regan:06a}.
Although potentially very powerful, the \lya method is  indirect and
hinges on a large number of assumptions (see e.g.~\cite{Boyarsky:08c}).

\paragraph{\lya bounds.}
\label{sec:lya-bounds-m_1}

A \lya analysis~\cite{Hansen:01,Viel:05,Viel:06,Seljak:06,Viel:07} has been
performed for NRP mechanism of sterile neutrinos, assuming a model with only
one sterile neutrino.  The bounds~\cite{Viel:05,Seljak:06,Viel:06} were
recently revised in Ref.~\cite{Boyarsky:08c}, who used the same SDSS \lya
data~\cite{McDonald:05} combined with WMAP5 data~\cite{Dunkley:2008ie}), and
paying special attention to the interpretation of statistics in the parameter
extraction and to possible systematic uncertainties. It was shown that a
conservative (i.e. frequentist, 3-$\sigma$) lower bound is $M_1>8$~keV. The
\lya method is still under development, and there is a possibility that some
of the related physical processes are not yet fully understood.  However, it
is difficult to identify a source of uncertainty that could give rise to
systematic errors affecting the result by more than $30\%$.  Even with such an
uncertainty, the possibility of all DM being in the form of NRP sterile
neutrinos can be ruled out by  comparing  \lya results with X-ray upper
bounds~\cite{Boyarsky:08c}.

A full \lya analysis of DM models containing both resonant and non-resonant
components has not yet been conducted. However in~\cite{Boyarsky:08c} a
cold-plus-warm DM model (CWDM) has been analyzed.  Although the phase-space
distribution of RP sterile neutrinos does not coincide exactly with such mixed
models, some results can be inferred from the CWDM analysis. In particular, it
has been shown~\cite{Boyarsky:08d}, that the cosmological signature of RP
sterile neutrino DM can be approximated by that of CWDM. With these
results~\cite{Boyarsky:08c}, it was shown that for each mass $M_1 \ge 2$~keV,
there exists at least one model of sterile neutrino that can account for the
totality of dark matter, and that is consistent with Lyman-$\alpha$ and other
cosmological data.  These corresponding values of lepton asymmetry can be
obtained within the \numsm.

\subsubsection{Galaxy substructures.}  

The suppression of the matter power spectrum at sub-Mpc scales may affect galaxy
substructures. Qualitatively one expects a cut-off scale for the existence of DM
substructures in the Milky Way--type galaxies. The primordial velocities of DM
particles do not allow them to move too close together, thereby creating shallow
density profiles.  Quantitative conclusions about the mass function of dwarf
satellites and slopes of density profiles of galactic substructures require
numerical simulations, as the galaxy formation today is at a non-linear stage.

A number of studies have claimed discrepancies between astrophysical
observations and CDM numerical simulations of galaxy formation (see
e.g.~\cite{Klypin:99,Ghigna:99,Moore:99,Sommer:99,Bode:00}). For example:
\begin{compactenum}[1.]
\item The number of observed dSphs of the Milky Way
  is still more than an order of magnitude below the CDM predictions, in spite
  of the drastically improved sensitivity in the dSphs search (see e.g.
  \cite{Gilmore:07a,Koposov:07});
\item direct measurements of DM density profiles in dSphs seem to favor cored
  profiles \cite{Kleyna:03,Goerdt:06,Sanchez:06b,Gilmore:07a};
\item there may be indications of the existence of the smallest scale at
  which DM is observed~\cite{Gilmore:07a,Strigari:08a}.
\end{compactenum}
However, it is not clear whether a ``CDM substructure crisis'' exists. Various
astrophysical resolutions of such a crisis has been proposed (see
e.g.~\cite{Kravtsov:98,Benson:01a,Strigari:06,Klimentowski:06,Strigari:06b,
  Penarrubia:07,Strigari:07,Koposov:09a}).  It is extremely challenging to
verify these resolutions both experimentally and numerically because the
full-scale numerical simulations including baryons and their feedback on
structure formation are not currently possible.  To understand how galaxy
structure formation occurs in the WDM models, one must perform numerical
simulations, that take into account thermal velocities of DM particles (see
e.g.~\cite{Colin:07},~\cite{Boyarsky:08c}).

\subsection{Particle physics experiments}
\label{sec:part-phys-exper}

Because many of the parameters of the \numsm are fixed or constrained by
experiments on neutrino masses and oscillations, as well as by cosmological
considerations, it is possible to rule out or to confirm this model.

We list below the \numsm predictions for experiments that are under way or
that will be carried out in the near future.

\begin{asparaenum}
\item Nothing but the Higgs boson within the mass interval $126{~\rm GeV} <M_H
  < 194{~\rm GeV}$ will be found at the
  LHC~\cite{Shaposhnikov:07b,Bezrukov:08a,Bezrukov:08b,Bezrukov:09a}
  (uncertainties in these numbers are given in~\cite{Bezrukov:09a}).

\item One of the active neutrinos in the \numsm must be very light, $m_1
  \lesssim {\cal O}(10^{-6})$~eV \cite{Asaka:05a,Boyarsky:06a}, which fixes
  the masses of the two other active neutrinos at $m_2\simeq 9\cdot 10^{-3}$
  eV and $m_3\simeq 5\cdot 10^{-2}$ eV for normal hierarchy or $m_{2,3}\simeq
  5 \cdot 10^{-2}$ eV for the inverted hierarchy. As a result, an effective
  Majorana mass for neutrinoless double-beta decay can be determined
  \cite{Bezrukov:2005mx}. For normal (inverted) hierarchy the constraints are
  $1.3~{\rm meV} <m_{\beta\beta}< 3.4~{\rm meV}$ ($13~{\rm meV}<
  m_{\beta\beta}< 50~{\rm meV}$).

\item 
  There will be negative results for the direct and indirect searches for
  weakly interacting massive particles, proton
  decay, neutron oscillations, and extra sources of CP-violation in the
  hadronic sector.
\end{asparaenum}
Although the experiments listed above can rule out the \numsm, they cannot
provide its direct confirmation. We can only obtain confirmation if all three
singlet fermions are discovered. Astrophysical searches for the DM sterile
neutrino are described at length above.  Laboratory searches for this particle
would require a challenging, detailed analysis of the kinematics of
$\beta$-decays of different isotopes \cite{Bezrukov:06}. A pair of new neutral
leptons, creating the baryon asymmetry of the universe, can be searched for
\cite{Gorbunov:07a} in dedicated experiments that use existing intensive
proton beams at CERN, FNAL, and planned neutrino facilities in Japan (J-PARC).
Alternatively, one can look for a specific missing energy signal in $K,~D$,
and $B$ decays. In general, the \numsm paradigm tells us that to uncover new
phenomena in particle physics we should utilize high-intensity proton beams
with fixed-target experiments or to very high intensity charm or B-factories,
rather than high energy accelerators.

\section{Conclusion}
\label{sec:conclusion}

Sterile neutrinos may play a key role in resolving the BSM problems.  For the specific
 choice of parameters described above, they can explain the neutrino
flavor oscillations and the origin of the matter-antimatter asymmetry in the
Universe, and can provide a DM candidate.
Sterile neutrino DM can be cold or warm, affecting the formation of structures
at sub-Mpc scales.  This parameter choice does not contradict  existing
experimental bounds and it assumes  masses below the electroweak scale. This
means that sterile neutrinos can be searched for in existing accelerator and
laboratory experiments. The parameter space of the sterile neutrinos in the
\numsm is bounded from all sides, allowing for verification or falsification
of this model.

\paragraph{Beyond the \numsm.}
\label{sec:beyond}

The \numsm is a minimal theory that can address the
SM problems in a unified way. Of course, the fact that it is minimal does not
mean that it is correct. Even if the low-energy electroweak theory indeed
contains three relatively light neutral right-handed leptonic states, the full
theory may be much more complicated. Extending the \numsm at low or high
energies will change the predictions of this model. For example, introducing
the low-energy supersymmetry would add another DM candidate to the theory,
allowing wider variation of the parameters of DM sterile neutrinos. If
GUT theories are proven to be true, they would introduce another
source of baryon number non-conservation, leading to a different origin of BAU.

Finally, we briefly discuss two modest modifications of the \numsm.  In the
first, one adds to the Lagrangian~(\ref{lagr1}) the higher-dimensional
operators given in Eq.~(\ref{lagr}). This allows for production of DM sterile
neutrinos at reheating after inflation \cite{Bezrukov:08a}, provided that they
are heavier than $100$ keV. This does not contradict to X-ray constraints
because the mixing angle $\theta_1$ is not essential for this production
mechanism. The sterile neutrino, in this case, can only play the role of CDM
candidate.

The second modification of the \numsm requires the addition of a singlet
scalar field \cite{Shaposhnikov:06}. This field may give a mass to singlet
fermions due to Yukawa couplings to these fermions, and it may play the role
of the inflaton~\cite{Shaposhnikov:06,Anisimov:08}. The decays of this scalar
to DM sterile neutrinos may serve as an efficient mechanism of cold or warm DM
production, without conflicting with X-ray
constraints~\cite{Shaposhnikov:06,Kusenko:06a,Petraki:07,Petraki:08,Boyanovsky:08b}.
In such models the mass of sterile neutrino DM can be in the MeV range or even
in the GeV range. The radiative decays of such DM particles can be searched
for with \emph{Fermi}~(cf.~\cite{Bertone:07}), whose large FoV requires a new
search strategy.

\paragraph{Astrophysical applications of sterile neutrinos.}
A final word about astrophysical applications of right-handed neutrinos,
discussed in this work. The mean free path of a sterile neutrino inside a
collapsing star is much larger than that of an ordinary neutrino. The escape
of sterile neutrinos from the supernova may thus provide an effective cooling
mechanism.  For sterile neutrinos with the masses $M\gtrsim 10^2$~keV the
analysis of the data from the supernova 1987A leads to an upper bound on the
mixing with electron neutrino $\sin^2(2\theta_e)\lesssim
10^{-10}$~\cite{Kainulainen:90,Raffelt:92,Raffelt-book,Raffelt:99}. In the
case of sterile neutrinos with the mass $M< 10^2$~keV the analysis shows that
the mixing with the electron neutrinos is bounded by $\sin^2(2\theta_e)
\lesssim 10^{-8} \left(\frac{\mathrm{keV}}{M}\right)^2$ for $1\kev\lesssim
M\lesssim 100\kev$~\cite{Raffelt:92,Shi:93}, while the mixing with
$\nu_{\mu,\tau}$ is bounded by $\sin^2(2\theta_{\mu,\tau}) \lesssim 3\times
10^{-8} T_{30}^{-6}$~\cite{Dolgov:00a} (c.f.~\cite{Abazajian:01a}) where
$T_{30} = T_{core}/30\mev$ and a typical supernovae core temperatures,
$T_{core}$, lie in the range $20-70\mev$~\cite{Prakash:96}.  Owing to large
uncertainties in the supernova physics, the constraints on the mixing angle of
the sterile neutrino that arise from these considerations happen to be weaker
than the X-ray bounds discussed above
(Section~\ref{sec:x-ray-constraints})~\cite{Dolgov:00a,Dolgov:00b}.

At the same time, the analysis of~\cite{Hidaka:06,Hidaka:07} shows that
sterile neutrinos with the masses $1\kev \lesssim M \lesssim 5\kev$ and with
the mixing angle with $\nu_e$ $10^{-10}\lesssim \sin^2(2\theta_e)\lesssim
10^{-8}$ may enhance lepton number, energy, and entropy transport in
supernovae explosions.  The results of~\cite{Fuller:08} indicate that sterile
neutrinos with the mass $M_{2,3}\sim 200\mev$ and mixing angles $\theta^2_\tau
> 10^{-8}$ or $10^{-8} < \theta_\mu^2 < 10^{-7}$ can augment shock energies of
core collapse supernovae.  The asymmetric emission of keV sterile neutrinos
from a supernova explosion may lead to an explanation of pulsar kick
velocities
\cite{Kusenko:1997sp,Barkovich:2004jp,Fuller:2003gy,Kusenko:06a,Petraki:07,Kusenko:08a},
see~\cite{Kusenko:09a} for a review.  The X-ray photons from early decays of
these particles can influence the formation of molecular hydrogen and can be
important for the early star formation and re-ionization
\cite{Biermann:06,Stasielak:06,Ripamonti:2006gr,
  Ripamonti:2006gq,Mapelli:2006ej}. See also
Refs.~\cite{Bilic:2001iv,Richter:2006fa,Munyaneza:06} for a discussion of the
formation of super-massive black holes and degenerate heavy neutrino stars.

Finally, if DM consists of many components (as may occur in extensions of the
\numsm) such that sterile neutrinos constitute only a fraction $f<1$ of it,
the X-ray bounds shown in Fig.~\ref{fig:sf-window} become weaker by the same
factor.  Moreover, the Lyman-$\alpha$ constraints also become diluted
\cite{Palazzo:07,Boyarsky:08c}, if the rest of DM is cold.  The analysis in
\cite{Boyarsky:08c} shows that the DM sterile neutrino with the spectrum given
by Eq.~(\ref{eq:19}) and a mass $M_1\ge 5$ keV for $f<0.6$ satisfies all
existing constraints at 99.7\% CL. Once $f<1$, sterile neutrino effects in
astrophysics may be even more important.

\subsection*{Acknowledgments}

O.R. and M.S.  acknowledge support of Swiss National Science Foundation.
\renewcommand{\bibsep}{0.5\baselineskip}

\bibliographystyle{nuke/arnuke_revised}

\begin{thebibliography}{100}

\bibitem{Glashow:1961tr}
S.~L. Glashow,
\newblock Nucl. Phys. {\bf 22}, 579 (1961).

\bibitem{Weinberg:1967tq}
S.~Weinberg,
\newblock Phys. Rev. Lett. {\bf 19}, 1264 (1967).

\bibitem{Salam:1968rm}
A.~Salam,
\newblock (1968),
\newblock Originally printed in *Svartholm: Elementary Particle Theory,
  Proceedings Of The Nobel Symposium Held 1968 At Lerum, Sweden*, Stockholm
  1968, 367-377.

\bibitem{Asaka:05a}
T.~Asaka, S.~Blanchet, and M.~Shaposhnikov,
\newblock Phys. Lett. {\bf B631}, 151 (2005), hep-ph/0503065.

\bibitem{Asaka:05b}
T.~{Asaka} and M.~{Shaposhnikov},
\newblock Phys. Lett. B {\bf 620}, 17 (2005), hep-ph/0505013.

\bibitem{Strumia:06}
A.~Strumia and F.~Vissani,
\newblock (2006), hep-ph/0606054.

\bibitem{Giunti:06}
C.~Giunti,
\newblock Nucl. Phys. Proc. Suppl. {\bf 169}, 309 (2007), hep-ph/0611125.

\bibitem{Schwetz:08a}
T.~Schwetz, M.~Tortola, and J.~W.~F. Valle,
\newblock New J. Phys. {\bf 10}, 113011 (2008), 0808.2016.

\bibitem{MINOS:08}
MINOS, P.~Adamson {\em et~al.},
\newblock Phys. Rev. Lett. {\bf 101}, 131802 (2008), 0806.2237.

\bibitem{KamLAND:08}
KamLAND, S.~Abe {\em et~al.},
\newblock Phys. Rev. Lett. {\bf 100}, 221803 (2008), 0801.4589.

\bibitem{Minkowski:77}
P.~Minkowski,
\newblock Phys. Lett. {\bf B67}, 421 (1977).

\bibitem{Ramond:79}
P.~Ramond,
\newblock (1979), hep-ph/9809459.

\bibitem{Mohapatra:79}
R.~N. Mohapatra and G.~Senjanovic,
\newblock Phys. Rev. Lett. {\bf 44}, 912 (1980).

\bibitem{Yanagida:80}
T.~Yanagida,
\newblock Prog. Theor. Phys. {\bf 64}, 1103 (1980).

\bibitem{Bezrukov:07}
F.~L. Bezrukov and M.~Shaposhnikov,
\newblock Phys. Lett. {\bf B659}, 703 (2008), 0710.3755.

\bibitem{Shaposhnikov:08b}
M.~Shaposhnikov and D.~Zenhausern,
\newblock Phys. Lett. {\bf B671}, 187 (2009), 0809.3395.

\bibitem{Shaposhnikov:08c}
M.~Shaposhnikov and D.~Zenhausern,
\newblock Phys. Lett. {\bf B671}, 162 (2009), 0809.3406.

\bibitem{Weidenspointner:06}
G.~{Weidenspointner} {\em et~al.},
\newblock \aap {\bf 450}, 1013 (2006), astro-ph/0601673.

\bibitem{Lyne:94}
A.~G. {Lyne} and D.~R. {Lorimer},
\newblock \nat {\bf 369}, 127 (1994).

\bibitem{PAMELA:08a}
PAMELA, O.~Adriani {\em et~al.},
\newblock Nature {\bf 458}, 607 (2009), 0810.4995.

\bibitem{nu0bb}
H.~V. Klapdor-Kleingrothaus, A.~Dietz, I.~V. Krivosheina, and O.~Chkvorets,
\newblock Nucl. Instrum. Meth. {\bf A522}, 371 (2004), hep-ph/0403018.

\bibitem{DAMA:08}
DAMA, R.~Bernabei {\em et~al.},
\newblock Eur. Phys. J. {\bf C56}, 333 (2008), 0804.2741.

\bibitem{Vissani:1997ys}
F.~Vissani,
\newblock Phys. Rev. {\bf D57}, 7027 (1998), hep-ph/9709409.

\bibitem{Fukugita:1986hr}
M.~Fukugita and T.~Yanagida,
\newblock Phys. Lett. {\bf B174}, 45 (1986).

\bibitem{Kuzmin:1985mm}
V.~A. Kuzmin, V.~A. Rubakov, and M.~E. Shaposhnikov,
\newblock Phys. Lett. {\bf B155}, 36 (1985).

\bibitem{Davidson:2008bu}
S.~Davidson, E.~Nardi, and Y.~Nir,
\newblock Phys. Rept. {\bf 466}, 105 (2008), 0802.2962.

\bibitem{deGouvea:05}
A.~de~Gouvea,
\newblock Phys. Rev. {\bf D72}, 033005 (2005), hep-ph/0501039.

\bibitem{Dodelson:93}
S.~Dodelson and L.~M. Widrow,
\newblock Phys. Rev. Lett. {\bf 72}, 17 (1994), hep-ph/9303287.

\bibitem{Shi:98}
X.-d. Shi and G.~M. Fuller,
\newblock Phys. Rev. Lett. {\bf 82}, 2832 (1999), astro-ph/9810076.

\bibitem{Dolgov:00}
A.~D. Dolgov and S.~H. Hansen,
\newblock Astropart. Phys. {\bf 16}, 339 (2002), hep-ph/0009083.

\bibitem{Abazajian:01a}
K.~Abazajian, G.~M. Fuller, and M.~Patel,
\newblock \prd {\bf 64}, 023501 (2001), astro-ph/0101524.

\bibitem{Abazajian:01b}
K.~Abazajian, G.~M. Fuller, and W.~H. Tucker,
\newblock \apj {\bf 562}, 593 (2001), astro-ph/0106002.

\bibitem{Tremaine:79}
S.~Tremaine and J.~E. Gunn,
\newblock Phys. Rev. Lett. {\bf 42}, 407 (1979).

\bibitem{Boyarsky:08a}
A.~Boyarsky, O.~Ruchayskiy, and D.~Iakubovskyi,
\newblock JCAP {\bf 0903}, 005 (2009), 0808.3902.

\bibitem{Pal:81}
P.~B. Pal and L.~Wolfenstein,
\newblock Phys. Rev. {\bf D25}, 766 (1982).

\bibitem{Barger:95}
V.~D. Barger, R.~J.~N. Phillips, and S.~Sarkar,
\newblock Phys. Lett. {\bf B352}, 365 (1995), hep-ph/9503295.

\bibitem{Boyarsky:08b}
A.~Boyarsky and O.~Ruchayskiy,
\newblock (2008), 0811.2385.

\bibitem{Asaka:06}
T.~Asaka, M.~Shaposhnikov, and A.~Kusenko,
\newblock Phys. Lett. {\bf B638}, 401 (2006), hep-ph/0602150.

\bibitem{Bezrukov:08b}
F.~L. Bezrukov, A.~Magnin, and M.~Shaposhnikov,
\newblock Phys. Lett. {\bf B675}, 88 (2009), 0812.4950.

\bibitem{Bezrukov:09a}
F.~Bezrukov and M.~Shaposhnikov,
\newblock (2009), 0904.1537.

\bibitem{DeSimone:2008ei}
A.~De~Simone, M.~P. Hertzberg, and F.~Wilczek,
\newblock Phys. Lett. {\bf B678}, 1 (2009), 0812.4946.

\bibitem{Barvinsky:2009fy}
A.~O. Barvinsky, A.~Y. Kamenshchik, C.~Kiefer, A.~A. Starobinsky, and
  C.~Steinwachs,
\newblock (2009), 0904.1698.

\bibitem{Bezrukov:08a}
F.~Bezrukov, D.~Gorbunov, and M.~Shaposhnikov,
\newblock JCAP {\bf 06}, 029 (2009), 0812.3622.

\bibitem{Dunkley:2008ie}
WMAP-5, J.~Dunkley {\em et~al.},
\newblock Astrophys. J. Suppl. {\bf 180}, 306 (2009), 0803.0586.

\bibitem{Sakharov:1967dj}
A.~D. Sakharov,
\newblock Pisma Zh. Eksp. Teor. Fiz. {\bf 5}, 32 (1967).

\bibitem{Kajantie:1996mn}
K.~Kajantie, M.~Laine, K.~Rummukainen, and M.~E. Shaposhnikov,
\newblock Phys. Rev. Lett. {\bf 77}, 2887 (1996), hep-ph/9605288.

\bibitem{Klinkhamer:1984di}
F.~R. Klinkhamer and N.~S. Manton,
\newblock Phys. Rev. {\bf D30}, 2212 (1984).

\bibitem{Shaposhnikov:08a}
M.~{Shaposhnikov},
\newblock JHEP {\bf 08}, 008 (2008), 0804.4542.

\bibitem{Akhmedov:98}
E.~K. Akhmedov, V.~A. Rubakov, and A.~Y. Smirnov,
\newblock Phys. Rev. Lett. {\bf 81}, 1359 (1998), hep-ph/9803255.

\bibitem{Dolgov:1980cq}
A.~D. Dolgov,
\newblock Sov. J. Nucl. Phys. {\bf 33}, 700 (1981).

\bibitem{Sigl:1992fn}
G.~Sigl and G.~Raffelt,
\newblock Nucl. Phys. {\bf B406}, 423 (1993).

\bibitem{Weldon:1982bn}
H.~A. Weldon,
\newblock Phys. Rev. {\bf D26}, 2789 (1982).

\bibitem{Laine:08a}
M.~{Laine} and M.~{Shaposhnikov},
\newblock JCAP {\bf 6}, 31 (2008), 0804.4543.

\bibitem{Serpico:05}
P.~D. Serpico and G.~G. Raffelt,
\newblock Phys. Rev. {\bf D71}, 127301 (2005), astro-ph/0506162.

\bibitem{Barbieri:1989ti}
R.~Barbieri and A.~Dolgov,
\newblock Phys. Lett. {\bf B237}, 440 (1990).

\bibitem{Notzold:87}
D.~Notzold and G.~Raffelt,
\newblock Nucl. Phys. {\bf B307}, 924 (1988).

\bibitem{Asaka:06b}
T.~Asaka, M.~Laine, and M.~Shaposhnikov,
\newblock JHEP {\bf 06}, 053 (2006), hep-ph/0605209.

\bibitem{Asaka:06c}
T.~Asaka, M.~Laine, and M.~Shaposhnikov,
\newblock JHEP {\bf 01}, 091 (2007), hep-ph/0612182.

\bibitem{Abazajian:02}
K.~N. Abazajian and G.~M. Fuller,
\newblock Phys. Rev. {\bf D66}, 023526 (2002), astro-ph/0204293.

\bibitem{Abazajian:05a}
K.~Abazajian,
\newblock Phys. Rev. {\bf D73}, 063506 (2006), astro-ph/0511630.

\bibitem{Boyanovsky:07a}
D.~Boyanovsky and C.~M. Ho,
\newblock Phys. Rev. {\bf D76}, 085011 (2007), 0705.0703.

\bibitem{Wolfenstein:1977ue}
L.~Wolfenstein,
\newblock Phys. Rev. {\bf D17}, 2369 (1978).

\bibitem{Mikheev:1986gs}
S.~P. Mikheev and A.~Y. Smirnov,
\newblock Sov. J. Nucl. Phys. {\bf 42}, 913 (1985).

\bibitem{Dolgov:00a}
A.~D. Dolgov, S.~H. Hansen, G.~Raffelt, and D.~V. Semikoz,
\newblock Nucl. Phys. {\bf B580}, 331 (2000), hep-ph/0002223.

\bibitem{Dolgov:00b}
A.~D. Dolgov, S.~H. Hansen, G.~Raffelt, and D.~V. Semikoz,
\newblock Nucl. Phys. {\bf B590}, 562 (2000), hep-ph/0008138.

\bibitem{Gorbunov:07a}
D.~Gorbunov and M.~Shaposhnikov,
\newblock JHEP {\bf 10}, 015 (2007), 0705.1729 [hep-ph].

\bibitem{Boyarsky:08c}
A.~Boyarsky, J.~Lesgourgues, O.~Ruchayskiy, and M.~Viel,
\newblock JCAP {\bf 0905}, 012 (2009), 0812.0010.

\bibitem{Peebles:80}
P.~J.~E. {Peebles},
\newblock {\em {The large-scale structure of the universe}} (Princeton, N.J.,
  Princeton University Press, 1980.~435 p., 1980).

\bibitem{Zeldovich:70}
Y.~B. {Zel'dovich},
\newblock \aap {\bf 5}, 84 (1970).

\bibitem{Bisnovatyi:80}
G.~S. {Bisnovatyi-Kogan},
\newblock \azh {\bf 57}, 899 (1980).

\bibitem{Bond:80}
J.~R. {Bond}, G.~{Efstathiou}, and J.~{Silk},
\newblock Phys. Rev. Lett. {\bf 45}, 1980 (1980).

\bibitem{Doroshkevich:81}
A.~G. {Doroshkevich}, M.~I. {Khlopov}, R.~A. {Sunyaev}, A.~S. {Szalay}, and
  I.~B. {Zeldovich},
\newblock New York Academy Sciences Annals {\bf 375}, 32 (1981).

\bibitem{Bode:00}
P.~Bode, J.~P. Ostriker, and N.~Turok,
\newblock \apj {\bf 556}, 93 (2001), astro-ph/0010389.

\bibitem{Viel:05}
M.~Viel, J.~Lesgourgues, M.~G. Haehnelt, S.~Matarrese, and A.~Riotto,
\newblock Phys. Rev. {\bf D71}, 063534 (2005), astro-ph/0501562.

\bibitem{Gorbunov:08b}
D.~Gorbunov, A.~Khmelnitsky, and V.~Rubakov,
\newblock JCAP {\bf 0810}, 041 (2008), 0808.3910.

\bibitem{Boyanovsky:08}
D.~{Boyanovsky}, H.~J. {de Vega}, and N.~G. {Sanchez},
\newblock \prd {\bf 77}, 043518 (2008), 0710.5180.

\bibitem{Boyarsky:06c}
A.~Boyarsky, A.~Neronov, O.~Ruchayskiy, M.~Shaposhnikov, and I.~Tkachev,
\newblock \prl {\bf 97}, 261302 (2006), astro-ph/0603660.

\bibitem{Boyarsky:06d}
A.~Boyarsky, J.~Nevalainen, and O.~Ruchayskiy,
\newblock \aap {\bf 471}, 51 (2007), astro-ph/0610961.

\bibitem{Boyarsky:06b}
A.~Boyarsky, A.~Neronov, O.~Ruchayskiy, and M.~Shaposhnikov,
\newblock \prd {\bf 74}, 103506 (2006), astro-ph/0603368.

\bibitem{Boyarsky:07a}
A.~{Boyarsky}, D.~{Iakubovskyi}, O.~{Ruchayskiy}, and V.~{Savchenko},
\newblock \mnras {\bf 387}, 1361 (2008), 0709.2301.

\bibitem{Boyarsky:05}
A.~Boyarsky, A.~Neronov, O.~Ruchayskiy, and M.~Shaposhnikov,
\newblock \mnras {\bf 370}, 213 (2006), astro-ph/0512509.

\bibitem{Abazajian:06b}
K.~N. {Abazajian}, M.~{Markevitch}, S.~M. {Koushiappas}, and R.~C. {Hickox},
\newblock \prd {\bf 75}, 063511 (2007), astro-ph/0611144.

\bibitem{Watson:06}
C.~R. Watson, J.~F. Beacom, H.~Yuksel, and T.~P. Walker,
\newblock Phys. Rev. {\bf D74}, 033009 (2006), astro-ph/0605424.

\bibitem{Boyarsky:06f}
A.~Boyarsky, J.~W. den Herder, A.~Neronov, and O.~Ruchayskiy,
\newblock Astropart.\ Phys. {\bf 28}, 303 (2007), astro-ph/0612219.

\bibitem{Riemer:06}
S.~{Riemer-S{\o}rensen}, S.~H. {Hansen}, and K.~{Pedersen},
\newblock \apjl {\bf 644}, L33 (2006), astro-ph/0603661.

\bibitem{Boyarsky:06e}
A.~Boyarsky, O.~Ruchayskiy, and M.~Markevitch,
\newblock \apj {\bf 673}, 752 (2008), astro-ph/0611168.

\bibitem{Yuksel:07}
H.~{Yuksel}, J.~F. {Beacom}, and C.~R. {Watson},
\newblock Phys. Rev. Lett. {\bf 101}, 121301 (2008), 0706.4084.

\bibitem{Boyarsky:07b}
A.~Boyarsky, D.~Malyshev, A.~Neronov, and O.~Ruchayskiy,
\newblock \mnras {\bf 387}, 1345 (2008), 0710.4922.

\bibitem{Loewenstein:08}
M.~Loewenstein, A.~Kusenko, and P.~L. Biermann,
\newblock (2008), 0812.2710.

\bibitem{Boyarsky:08d}
A.~Boyarsky, J.~Lesgourgues, O.~Ruchayskiy, and M.~Viel,
\newblock Phys. Rev. Lett. {\bf 102}, 201304 (2009), 0812.3256.

\bibitem{denHerder:09}
J.~W. den Herder, A.~Boyarsky, O.~Ruchayskiy, {\em et~al.},
\newblock (2009), 0906.1788.

\bibitem{Piro:08}
L.~{Piro} {\em et~al.},
\newblock Nuovo Cim. {\bf 122B}, 1007 (2007), 0707.4103.

\bibitem{Bi:93}
H.~{Bi},
\newblock \apj {\bf 405}, 479 (1993).

\bibitem{Viel:2001hd}
M.~Viel, S.~Matarrese, H.~J. Mo, M.~G. Haehnelt, and T.~Theuns,
\newblock Mon. Not. Roy. Astron. Soc. {\bf 329}, 848 (2002), astro-ph/0105233.

\bibitem{Zaldarriaga:2001xs}
M.~Zaldarriaga, R.~Scoccimarro, and L.~Hui,
\newblock Astrophys. J. {\bf 590}, 1 (2003), astro-ph/0111230.

\bibitem{Lewis:99}
A.~Lewis, A.~Challinor, and A.~Lasenby,
\newblock Astrophys. J. {\bf 538}, 473 (2000), astro-ph/9911177.

\bibitem{Theuns:98}
T.~Theuns, A.~Leonard, G.~Efstathiou, F.~R. Pearce, and P.~A. Thomas,
\newblock Mon. Not. Roy. Astron. Soc. {\bf 301}, 478 (1998), astro-ph/9805119.

\bibitem{Gnedin:01}
N.~Y. Gnedin and A.~J.~S. Hamilton,
\newblock Mon.Not.Roy.Astron.Soc. {\bf 334}, 107 (2002), astro-ph/0111194.

\bibitem{McDonald:05}
P.~{McDonald} {\em et~al.},
\newblock \apjs {\bf 163}, 80 (2006), astro-ph/0405013.

\bibitem{Viel:04}
M.~Viel, M.~G. Haehnelt, and V.~Springel,
\newblock Mon. Not. Roy. Astron. Soc. {\bf 354}, 684 (2004), astro-ph/0404600.

\bibitem{Viel:05b}
M.~Viel, M.~G. Haehnelt, and V.~Springel,
\newblock Mon.Not.Roy.Astron.Soc. {\bf 367}, 1655 (2006), astro-ph/0504641.

\bibitem{Viel:05c}
M.~Viel and M.~G. Haehnelt,
\newblock Mon. Not. Roy. Astron. Soc. {\bf 365}, 231 (2006), astro-ph/0508177.

\bibitem{Regan:06a}
J.~A. Regan, M.~G. Haehnelt, and M.~Viel,
\newblock Mon. Not. Roy. Astron. Soc. {\bf 374}, 196 (2007), astro-ph/0606638.

\bibitem{Hansen:01}
S.~H. Hansen, J.~Lesgourgues, S.~Pastor, and J.~Silk,
\newblock \mnras {\bf 333}, 544 (2002), astro-ph/0106108.

\bibitem{Viel:06}
M.~Viel, J.~Lesgourgues, M.~G. Haehnelt, S.~Matarrese, and A.~Riotto,
\newblock Phys. Rev. Lett. {\bf 97}, 071301 (2006), astro-ph/0605706.

\bibitem{Seljak:06}
U.~Seljak, A.~Makarov, P.~McDonald, and H.~Trac,
\newblock Phys. Rev. Lett. {\bf 97}, 191303 (2006), astro-ph/0602430.

\bibitem{Viel:07}
M.~{Viel} {\em et~al.},
\newblock Phys. Rev. Lett. {\bf 100}, 041304 (2008), 0709.0131.

\bibitem{Klypin:99}
A.~{Klypin}, A.~V. {Kravtsov}, O.~{Valenzuela}, and F.~{Prada},
\newblock \apj {\bf 522}, 82 (1999), astro-ph/9901240.

\bibitem{Ghigna:99}
S.~{Ghigna} {\em et~al.},
\newblock \apj {\bf 544}, 616 (2000), astro-ph/9910166.

\bibitem{Moore:99}
B.~{Moore}, T.~{Quinn}, F.~{Governato}, J.~{Stadel}, and G.~{Lake},
\newblock \mnras {\bf 310}, 1147 (1999), astro-ph/9903164.

\bibitem{Sommer:99}
J.~{Sommer-Larsen} and A.~{Dolgov},
\newblock \apj {\bf 551}, 608 (2001), astro-ph/9912166.

\bibitem{Gilmore:07a}
G.~{Gilmore} {\em et~al.},
\newblock \apj {\bf 663}, 948 (2007), astro-ph/0703308.

\bibitem{Koposov:07}
S.~{Koposov} {\em et~al.},
\newblock \apj {\bf 663}, 948 (2007), 0706.2687.

\bibitem{Kleyna:03}
J.~T. {Kleyna}, M.~I. {Wilkinson}, G.~{Gilmore}, and N.~W. {Evans},
\newblock \apjl {\bf 588}, L21 (2003), astro-ph/0304093.

\bibitem{Goerdt:06}
T.~Goerdt, B.~Moore, J.~I. Read, J.~Stadel, and M.~Zemp,
\newblock \mnras {\bf 368}, 1073 (2006), astro-ph/0601404.

\bibitem{Sanchez:06b}
F.~J. {S{\'a}nchez-Salcedo}, J.~{Reyes-Iturbide}, and X.~{Hernandez},
\newblock \mnras {\bf 370}, 1829 (2006), astro-ph/0601490.

\bibitem{Strigari:08a}
L.~E. Strigari {\em et~al.},
\newblock Nature {\bf 454}, 1096 (2008), 0808.3772.

\bibitem{Kravtsov:98}
A.~V. {Kravtsov}, A.~A. {Klypin}, J.~S. {Bullock}, and J.~R. {Primack},
\newblock \apj {\bf 502}, 48 (1998), astro-ph/9708176.

\bibitem{Benson:01a}
A.~J. {Benson}, C.~G. {Lacey}, C.~M. {Baugh}, S.~{Cole}, and C.~S. {Frenk},
\newblock \mnras {\bf 333}, 156 (2002), astro-ph/0108217.

\bibitem{Strigari:06}
L.~E. Strigari {\em et~al.},
\newblock \apj {\bf 652}, 306 (2006), astro-ph/0603775.

\bibitem{Klimentowski:06}
J.~{Klimentowski} {\em et~al.},
\newblock \mnras {\bf 378}, 353 (2007), astro-ph/0611296.

\bibitem{Strigari:06b}
L.~E. Strigari, M.~Kaplinghat, and J.~S. Bullock,
\newblock Phys. Rev. {\bf D75}, 061303 (2007), astro-ph/0606281.

\bibitem{Penarrubia:07}
J.~{Penarrubia}, A.~{McConnachie}, and J.~F. {Navarro},
\newblock \apj {\bf 672}, 904 (2008), astro-ph/0701780.

\bibitem{Strigari:07}
L.~E. {Strigari} {\em et~al.},
\newblock \apj {\bf 669}, 676 (2007), 0704.1817.

\bibitem{Koposov:09a}
S.~E. {Koposov} {\em et~al.},
\newblock (2009), 0901.2116.

\bibitem{Colin:07}
P.~Colin, O.~Valenzuela, and V.~Avila-Reese,
\newblock Astrophys. J. {\bf 673}, 203 (2008), 0709.4027.

\bibitem{Shaposhnikov:07b}
M.~Shaposhnikov,
\newblock (2007), 0708.3550.

\bibitem{Boyarsky:06a}
A.~Boyarsky, A.~Neronov, O.~Ruchayskiy, and M.~Shaposhnikov,
\newblock JETP Letters , 133 (2006), hep-ph/0601098.

\bibitem{Bezrukov:2005mx}
F.~Bezrukov,
\newblock Phys. Rev. {\bf D72}, 071303 (2005), hep-ph/0505247.

\bibitem{Bezrukov:06}
F.~{Bezrukov} and M.~{Shaposhnikov},
\newblock \prd {\bf 75}, 053005 (2007), hep-ph/0611352.

\bibitem{Shaposhnikov:06}
M.~Shaposhnikov and I.~Tkachev,
\newblock Phys. Lett. {\bf B639}, 414 (2006), hep-ph/0604236.

\bibitem{Anisimov:08}
A.~Anisimov, Y.~Bartocci, and F.~L. Bezrukov,
\newblock Phys. Lett. {\bf B671}, 211 (2009), 0809.1097.

\bibitem{Kusenko:06a}
A.~Kusenko,
\newblock Phys. Rev. Lett. {\bf 97}, 241301 (2006), hep-ph/0609081.

\bibitem{Petraki:07}
K.~Petraki and A.~Kusenko,
\newblock Phys. Rev. {\bf D77}, 065014 (2008), 0711.4646.

\bibitem{Petraki:08}
K.~{Petraki},
\newblock \prd {\bf 77}, 105004 (2008), 0801.3470.

\bibitem{Boyanovsky:08b}
D.~Boyanovsky,
\newblock Phys. Rev. {\bf D78}, 103505 (2008), 0807.0646.

\bibitem{Bertone:07}
G.~Bertone, W.~Buchmuller, L.~Covi, and A.~Ibarra,
\newblock JCAP {\bf 0711}, 003 (2007), 0709.2299.

\bibitem{Kainulainen:90}
K.~Kainulainen, J.~Maalampi, and J.~T. Peltoniemi,
\newblock Nucl. Phys. {\bf B358}, 435 (1991).

\bibitem{Raffelt:92}
G.~Raffelt and G.~Sigl,
\newblock Astropart. Phys. {\bf 1}, 165 (1993), astro-ph/9209005.

\bibitem{Raffelt-book}
G.~G. Raffelt,
\newblock {\em Stars as laboratories for fundamental physics: The astrophysics
  of neutrinos, axions, and other weakly interacting particles} (University of
  Chicago Press, Chicago, USA, 1996).

\bibitem{Raffelt:99}
G.~G. Raffelt,
\newblock Ann. Rev. Nucl. Part. Sci. {\bf 49}, 163 (1999), hep-ph/9903472.

\bibitem{Shi:93}
X.~Shi and G.~Sigl,
\newblock Phys. Lett. {\bf B323}, 360 (1994), hep-ph/9312247.

\bibitem{Prakash:96}
M.~Prakash {\em et~al.},
\newblock Phys. Rept. {\bf 280}, 1 (1997), nucl-th/9603042.

\bibitem{Hidaka:06}
J.~Hidaka and G.~M. Fuller,
\newblock \prd {\bf 74}, 125015 (2006), astro-ph/0609425.

\bibitem{Hidaka:07}
J.~{Hidaka} and G.~M. {Fuller},
\newblock \prd {\bf 76}, 083516 (2007), 0706.3886.

\bibitem{Fuller:08}
G.~M. Fuller, A.~Kusenko, and K.~Petraki,
\newblock Phys. Lett. {\bf B670}, 281 (2009), 0806.4273.

\bibitem{Kusenko:1997sp}
A.~Kusenko and G.~Segre,
\newblock Phys. Lett. {\bf B396}, 197 (1997), hep-ph/9701311.

\bibitem{Barkovich:2004jp}
M.~Barkovich, J.~C. D'Olivo, and R.~Montemayor,
\newblock Phys. Rev. {\bf D70}, 043005 (2004), hep-ph/0402259.

\bibitem{Fuller:2003gy}
G.~M. Fuller, A.~Kusenko, I.~Mocioiu, and S.~Pascoli,
\newblock Phys. Rev. {\bf D68}, 103002 (2003), astro-ph/0307267.

\bibitem{Kusenko:08a}
A.~Kusenko, B.~P. Mandal, and A.~Mukherjee,
\newblock Phys. Rev. {\bf D77}, 123009 (2008), 0801.4734.

\bibitem{Kusenko:09a}
A.~Kusenko,
\newblock (2009), 0906.2968.

\bibitem{Biermann:06}
P.~L. Biermann and A.~Kusenko,
\newblock Phys. Rev. Lett. {\bf 96}, 091301 (2006), astro-ph/0601004.

\bibitem{Stasielak:06}
J.~Stasielak, P.~L. Biermann, and A.~Kusenko,
\newblock \apj {\bf 654}, 290 (2007), astro-ph/0606435.

\bibitem{Ripamonti:2006gr}
E.~Ripamonti, M.~Mapelli, and A.~Ferrara,
\newblock Mon. Not. Roy. Astron. Soc. {\bf 375}, 1399 (2007), astro-ph/0606483.

\bibitem{Ripamonti:2006gq}
E.~Ripamonti, M.~Mapelli, and A.~Ferrara,
\newblock Mon. Not. Roy. Astron. Soc. {\bf 374}, 1067 (2007), astro-ph/0606482.

\bibitem{Mapelli:2006ej}
M.~Mapelli, A.~Ferrara, and E.~Pierpaoli,
\newblock Mon. Not. Roy. Astron. Soc. {\bf 369}, 1719 (2006), astro-ph/0603237.

\bibitem{Bilic:2001iv}
N.~Bilic, R.~J. Lindebaum, G.~B. Tupper, and R.~D. Viollier,
\newblock Phys. Lett. {\bf B515}, 105 (2001), astro-ph/0106209.

\bibitem{Richter:2006fa}
M.~C. Richter, G.~B. Tupper, and R.~D. Viollier,
\newblock JCAP {\bf 0612}, 015 (2006), astro-ph/0611552.

\bibitem{Munyaneza:06}
F.~{Munyaneza} and P.~L. {Biermann},
\newblock \aap {\bf 458}, L9 (2006), astro-ph/0609388.

\bibitem{Palazzo:07}
A.~Palazzo, D.~Cumberbatch, A.~Slosar, and J.~Silk,
\newblock Phys. Rev. {\bf D76}, 103511 (2007), 0707.1495.

\end{thebibliography}

\let\jnlstyle=\rm\def\jref#1{{\jnlstyle#1}}\def\aj{\jref{AJ}}
  \def\araa{\jref{ARA\&A}} \def\apj{\jref{ApJ}\ } \def\apjl{\jref{ApJ}\ }
  \def\apjs{\jref{ApJS}} \def\ao{\jref{Appl.~Opt.}} \def\apss{\jref{Ap\&SS}}
  \def\aap{\jref{A\&A}} \def\aapr{\jref{A\&A~Rev.}} \def\aaps{\jref{A\&AS}}
  \def\azh{\jref{AZh}} \def\baas{\jref{BAAS}} \def\jrasc{\jref{JRASC}}
  \def\memras{\jref{MmRAS}} \def\mnras{\jref{MNRAS}\ }
  \def\pra{\jref{Phys.~Rev.~A}\ } \def\prb{\jref{Phys.~Rev.~B}\ }
  \def\prc{\jref{Phys.~Rev.~C}\ } \def\prd{\jref{Phys.~Rev.~D}\ }
  \def\pre{\jref{Phys.~Rev.~E}} \def\prl{\jref{Phys.~Rev.~Lett.}}
  \def\pasp{\jref{PASP}} \def\pasj{\jref{PASJ}} \def\qjras{\jref{QJRAS}}
  \def\skytel{\jref{S\&T}} \def\solphys{\jref{Sol.~Phys.}}
  \def\sovast{\jref{Soviet~Ast.}} \def\ssr{\jref{Space~Sci.~Rev.}}
  \def\zap{\jref{ZAp}} \def\nat{\jref{Nature}\ } \def\iaucirc{\jref{IAU~Circ.}}
  \def\aplett{\jref{Astrophys.~Lett.}}
  \def\apspr{\jref{Astrophys.~Space~Phys.~Res.}}
  \def\bain{\jref{Bull.~Astron.~Inst.~Netherlands}}
  \def\fcp{\jref{Fund.~Cosmic~Phys.}} \def\gca{\jref{Geochim.~Cosmochim.~Acta}}
  \def\grl{\jref{Geophys.~Res.~Lett.}} \def\jcp{\jref{J.~Chem.~Phys.}}
  \def\jgr{\jref{J.~Geophys.~Res.}}
  \def\jqsrt{\jref{J.~Quant.~Spec.~Radiat.~Transf.}}
  \def\memsai{\jref{Mem.~Soc.~Astron.~Italiana}}
  \def\nphysa{\jref{Nucl.~Phys.~A}} \def\physrep{\jref{Phys.~Rep.}}
  \def\physscr{\jref{Phys.~Scr}} \def\planss{\jref{Planet.~Space~Sci.}}
  \def\procspie{\jref{Proc.~SPIE}} \let\astap=\aap \let\apjlett=\apjl
  \let\apjsupp=\apjs \let\applopt=\ao

\end{document}